\documentclass[prb,twocolumn,superscriptaddress,nopacs]{revtex4}
\usepackage{graphicx}

\usepackage{amsmath}
\usepackage{amssymb}
\usepackage{accents}
\usepackage{color}
\usepackage{verbatim}
\usepackage{hyperref}
\usepackage{empheq} 
\definecolor{very-light-gray}{gray}{0.9}

\usepackage{float} 

\newcommand{\bk}{{\bf k}}

\renewcommand{\r}{\mathbf{r}}

\begin{document}

\title{Charge-transfer insulation in twisted bilayer graphene}
\author{Louk Rademaker}
\affiliation{Department of Theoretical Physics, University of Geneva, 1211 Geneva, Switzerland}
\affiliation{Perimeter Institute for Theoretical Physics, Waterloo, Ontario N2L 2Y5, Canada}
\author{Paula Mellado}
\affiliation{Perimeter Institute for Theoretical Physics, Waterloo, Ontario N2L 2Y5, Canada}
\affiliation{School of Engineering and Sciences, Adolfo Ib\'{a}\~{n}ez University, Santiago 7941169, Chile}

\date{\today}

\begin{abstract} 
We studied the real space structure of states in twisted bilayer graphene at the `magic angle' $\theta = 1.08^\circ$. The flat bands close to charge neutrality are composed of a mix of `ring' and `center' orbitals around the AA stacking region. An effective model with localized orbitals is constructed, which necessarily includes more than just the four flat bands. Long-range Coulomb interaction causes a charge-transfer at half-filling of the flat bands from the `center' to the `ring' orbitals. Consequently, the Mott phase is a featureless spin-singlet paramagnet. We estimate the effective Heisenberg coupling that favors the singlet coupling to be $J = 3.3$ K, consistent with experimental values. The superconducting state depends on the nature of the dopants: hole-doping yields $p+ip$-wave whereas electron-doping yields $d+id$-wave pairing symmetry.
\end{abstract}


\maketitle

\begin{figure}[H]
	\includegraphics[width=0.24\columnwidth]{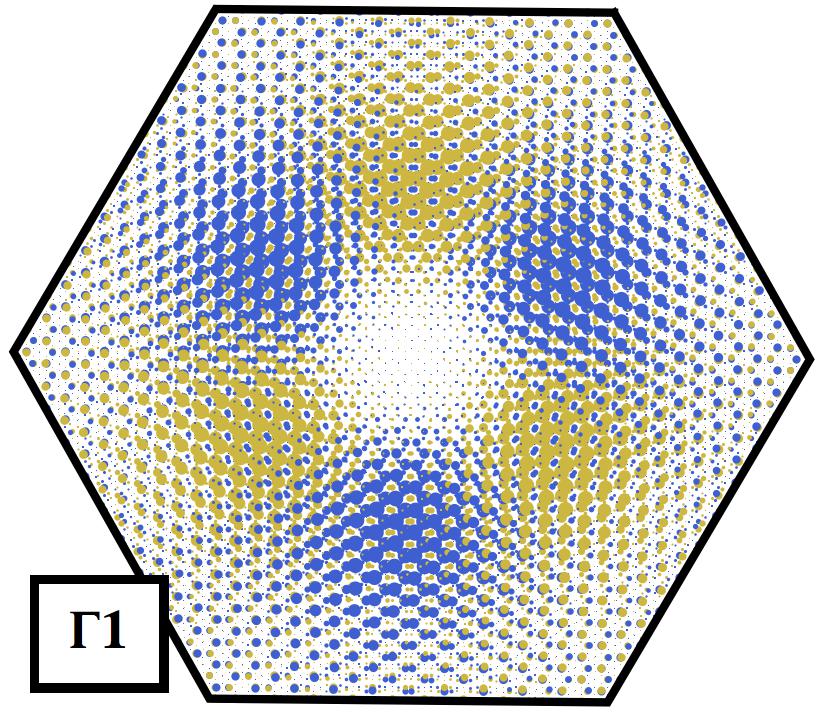}
	\includegraphics[width=0.24\columnwidth]{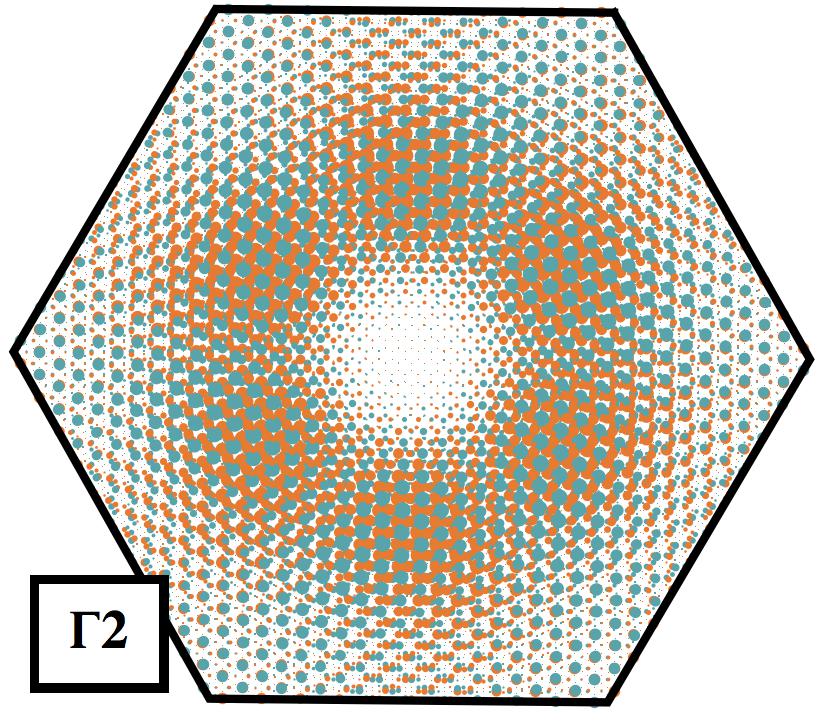}
	\includegraphics[width=0.24\columnwidth]{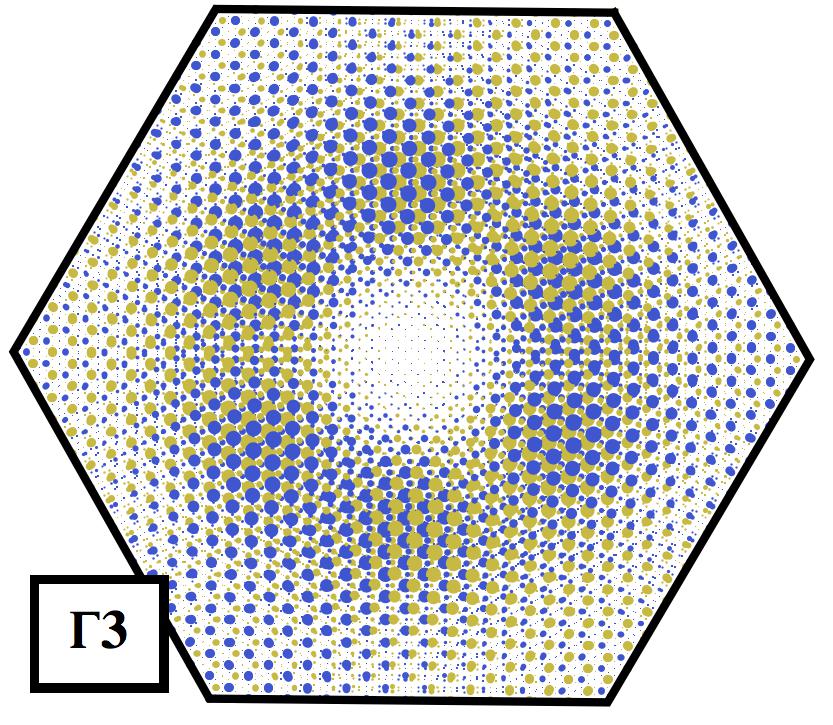}
	\includegraphics[width=0.24\columnwidth]{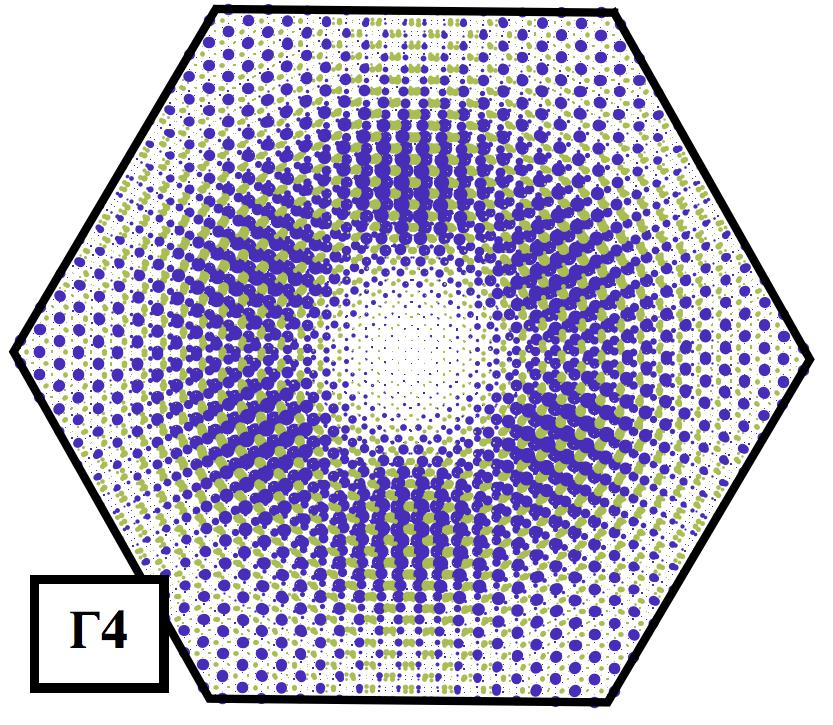}
	\includegraphics[width=0.24\columnwidth]{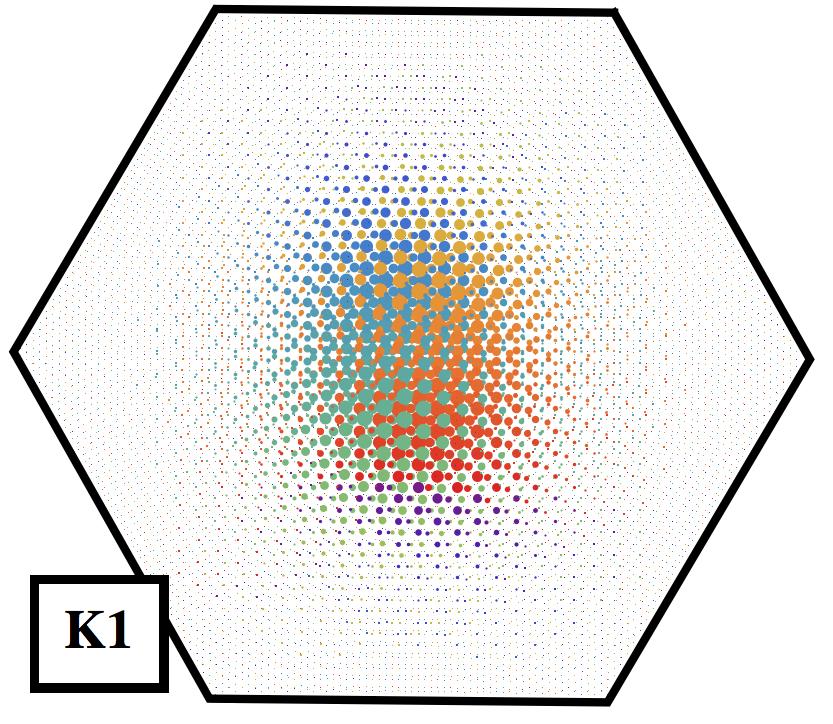}
	\includegraphics[width=0.24\columnwidth]{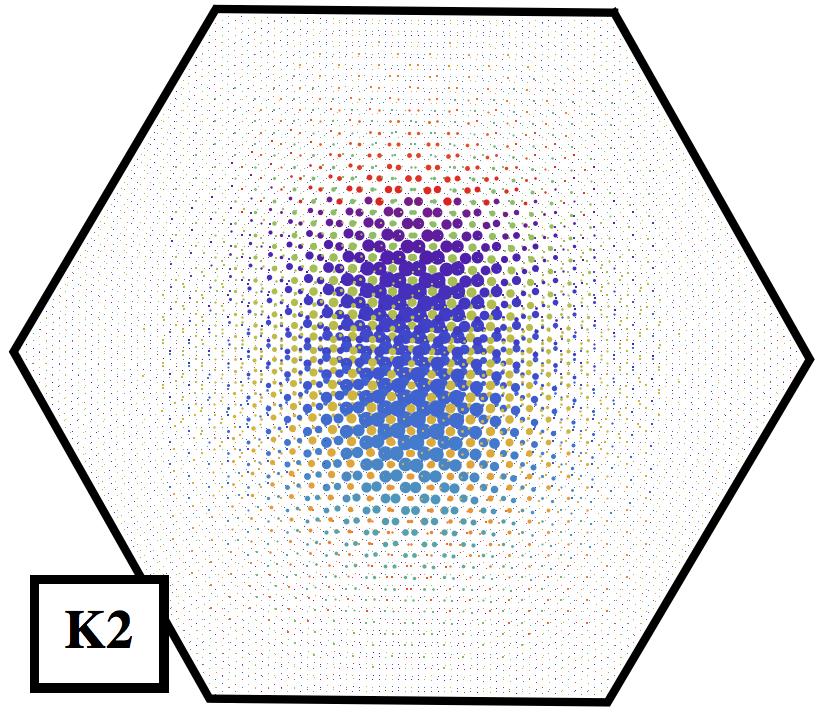}
	\includegraphics[width=0.24\columnwidth]{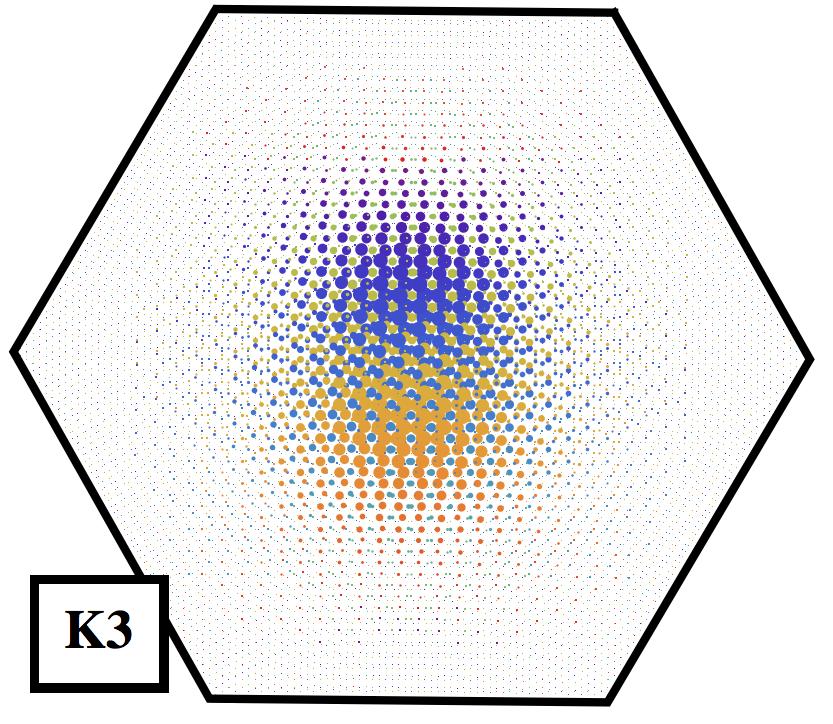}	
	\includegraphics[width=0.24\columnwidth]{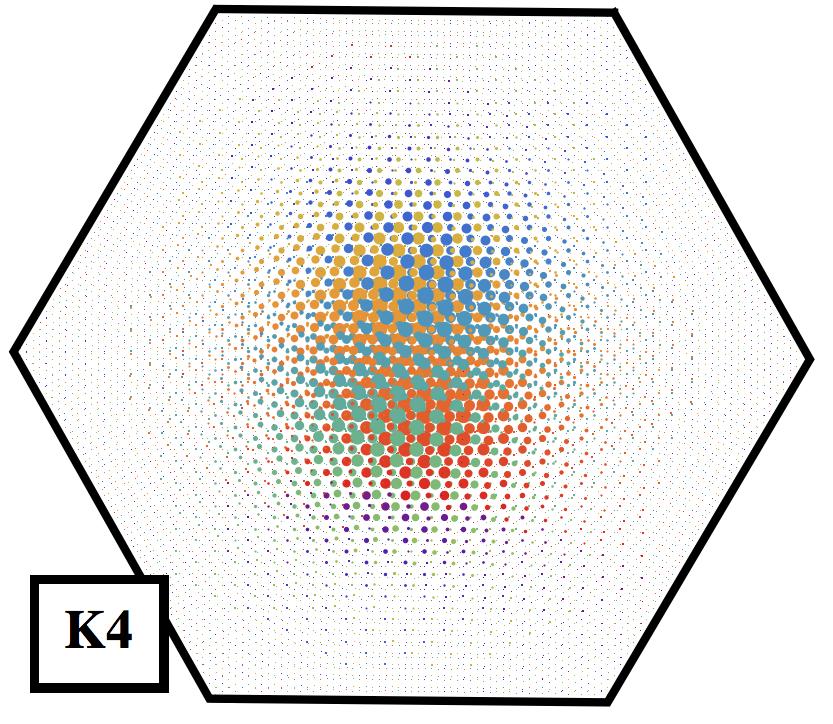}
	\includegraphics[width=\columnwidth]{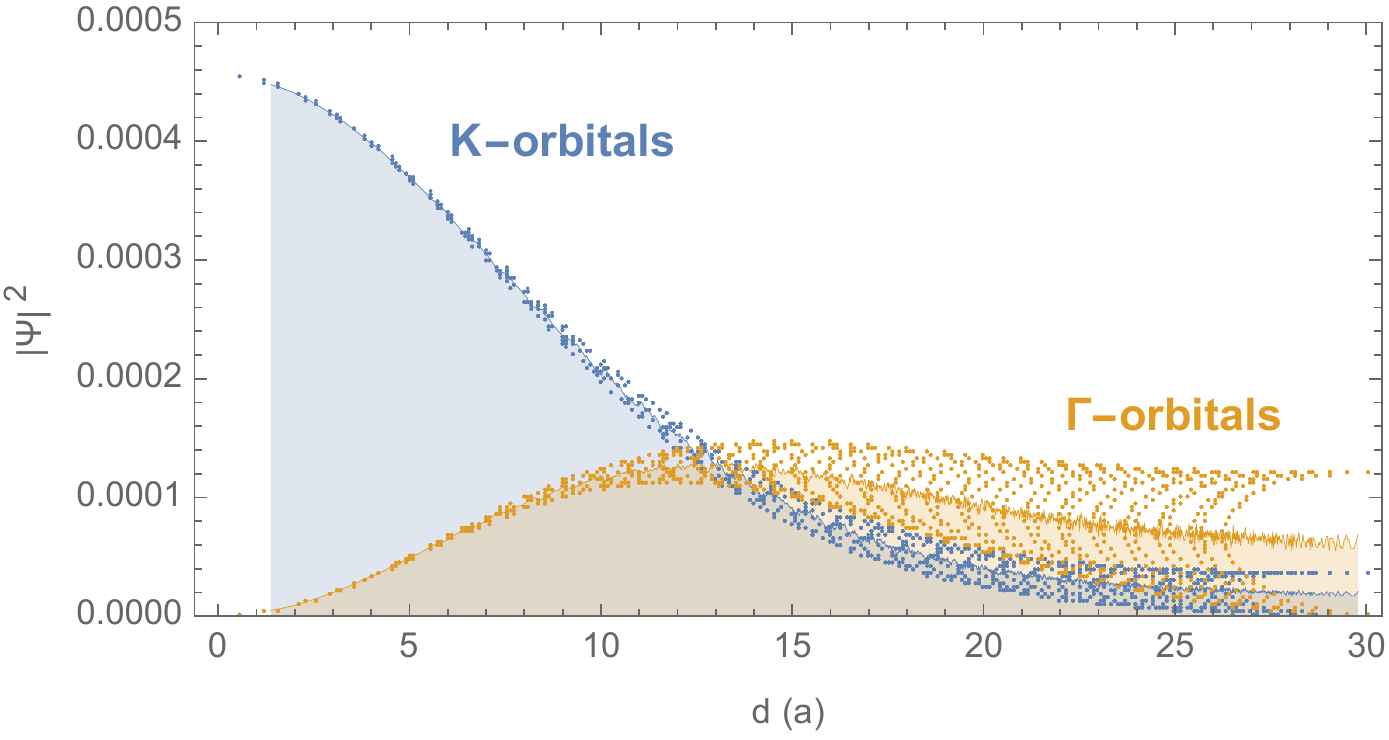}
	\caption{\label{FigOrbitals}
	The wavefunctions of the lowest energy orbitals at ${\bf \Gamma}$ (top row) and ${\bf K}$ (middle row), computed using the tight-binding model of Fig. \ref{FigFlatBands}. Shown here is the hexagonal unit cell with the AA stacking at the center, and the six corners having alternating AB and BA stacking. The size of the dots is proportional to the wavefunction modulus squared on each atom, the color represents the phase. 
	The ${\bf K}$-orbitals have large spectral weight at the center of the AA region.
	The ${\bf \Gamma}$-orbitals, on the other hand, have suppressed weight at the AA center but have nonzero weight in a ring around it. 
	The averaged wavefunction-squared as a function of distance from the AA centers, measured in units of the single layer graphene lattice parameter $a$, is shown in the bottom plot. We propose that the insulating phase is characterized by a charge-transfer from the `center' to the `ring'-orbitals due to long-range Coulomb repulsion. Note that there is a nonzero overlap between the ${\bf \Gamma}$ and ${\bf K}$-orbitals. The orbital nature of the flat band smoothly varies from `ring' to `center' orbitals as a function of momentum.
	The {\em qualitative} difference between the real-space wavefunction of the flat bands at ${\bf \Gamma}$ and ${\bf K}$ implies that one need more than four {\em localized} Wannier orbitals to capture the flat bands correctly. 
	}
\end{figure}

\section{Introduction}

Upon the discovery of superconductivity in twisted bilayer graphene (TBG)\cite{Cao:2018kn,Cao:2018ff,Yankowitz:2018tx}, the theoretical community jumped on the `flat band'-wagon and started the hunt for a simple yet sufficient model that might shed light on the nature and symmetries of this system\cite{Efimkin:2018us,Pal:2018wc,Xu:2018,Yuan:2018un,Po:2018,Roy:2018,Guo:2018,Padhi:2018,Irkhin:2018,Dodaro:2018,Zhang:2018,Zhang:2018wj,Thomson:2018ve,Su:2018vz,Zou:2018ud,Pizarro:2018wx,Isobe:2018wx,Wu:2018vh,You:2018ud,Sboychakov:2018tp,Xu:2018vb,Ochi:2018ug,Sherkunov:2018wf,Kennes:2018wi,Koshino:2018td,Kang:2018wd,Fidrysiak:2018,Song:2018ul,Venderbos:2018vs,Po:2018vk,Tang:2018tw,Laksono:2018dm,Tarnopolsky:2018vv,Chen:2018vz,Lin:2018tx,Fu:2018vx,Marzari:2012eu,Marzari:1997wa}. The fact that the phase diagram, when squinting, looks similar to that of cuprates or pnictides made this discovery even more exhilarating. Namely, at half-filling of the double degenerate flat band just below the Dirac cones a Mott insulator was found, and superconductivity appeared upon doping of this Mott state.


The nature of a Mott insulating phase, especially when dealing with degenerate bands, can only be fully understood by studying the {\em real-space structure} of the orbitals that are involved. At charge-neutrality, the electronic density of states is peaked around the `AA' centers of the large unit cell. This observation is at the core of practically all theoretical work on TBG since March, by constructing an effective model either on the triangular lattice of the AA centers\cite{Xu:2018,Guo:2018,Dodaro:2018,Wu:2018vh,You:2018ud,Fidrysiak:2018,Thomson:2018ve}, or on the honeycomb lattice of AB/BA centers with so-called 'fidget spinner' orbitals that are peaked at three neighboring AA centers\cite{Po:2018,Zou:2018ud,Koshino:2018td,Ochi:2018ug,Kang:2018wd,Yuan:2018un}. However, a crucial aspect is missed, namely that {\em the real-space structure of the flat bands qualitatively changes when doping away from charge neutrality}. This implies that an effective model with {\em localized} Wannier orbitals needs more than just the four flat bands.

This real-space change is shown in Fig. \ref{FigOrbitals}, where we compared the real-space structure of the flat band states at charge neutrality (the ${\bf K}$-point) to the states at the other end of the flat bands (the ${\bf \Gamma}$-point). In agreement with previous works,\cite{Wong:2015dg,Cao:2018kn} we also find  that at the ${\bf K}$-point charge is localized at the AA centers. However, at ${\bf \Gamma}$ the charge density vanishes at the AA points - it is moved to a ring around it. The usual construction of maximally localized Wannier orbitals\cite{Marzari:1997wa,Marzari:2012eu} will yield necessarily orbitals that will be larger than the unit cell - the 'fidget spinner' orbitals are a prime example.\cite{Po:2018,Zou:2018ud,Po:2018vk,Kang:2018wd} 

At this point, the reader might wonder why we should care about this change. The answer is simple: because we are interested in the Mott insulating state. The real-space structure of orbitals directly determines the relevant Hubbard and exchange couplings of the model. Furthermore, Wannier orbitals that extend over several unit cells are incompatible with the physics of Mott localization, which requires localized orbitals! If one insists on a triangular lattice model, then an on-site SU(4) symmetric Hubbard coupling is natural\cite{Xu:2018}. If one chooses the fidget spinner orbitals, then nearest-neighbor and next-nearest-neighbor repulsion will play an important role\cite{Koshino:2018td}. However, both these models neglect the fact that the flat bands change orbital character in between the ${\bf K}$ and ${\bf \Gamma}$ points. In this paper we choose {\em not} to neglect this change, and we will show that it has important consequences. In particular, we will show that the interaction energy is minimized in the insulating phase by a charge-transfer from the `center' to the `ring' orbitals. The remaining spin degrees of freedom are subject to an effective Heisenberg coupling, forcing them into an onsite singlet. The Heisenberg coupling has a strength of the order of $J \sim 3.3$ K, consistent with the experimental energy scale that destroys the insulating state by either temperature or magnetic field\cite{Cao:2018kn,Cao:2018ff}.


This paper is organized as follows: In section \ref{model} we discuss the symmetries and structure of the twisted bilayer graphene lattice. We then construct the real-space wavefunctions and a corresponding effective tight binding model. Using the hopping model and general energy considerations, in section \ref{interactions} we study the insulating phase and show that the interaction energy is minimized in the insulating phase by a charge-transfer from the `center' to the `ring' orbital." The transfer of charge picture permits to fix the spin degrees of freedom in the Mott phase yielding naturally an effective antiferromagnetic Heisenberg hamiltonian which gives rise to a spin singlet paramagnet. In section \ref{superconductivity} we examine how doping this paramagnet can lead to superconductivity, and the different pairing symmetries for different doping are discussed. We end the article with a brief discussion of experimental signatures of the suggested charge-transfer.




\begin{figure}
	\includegraphics[width=\columnwidth]{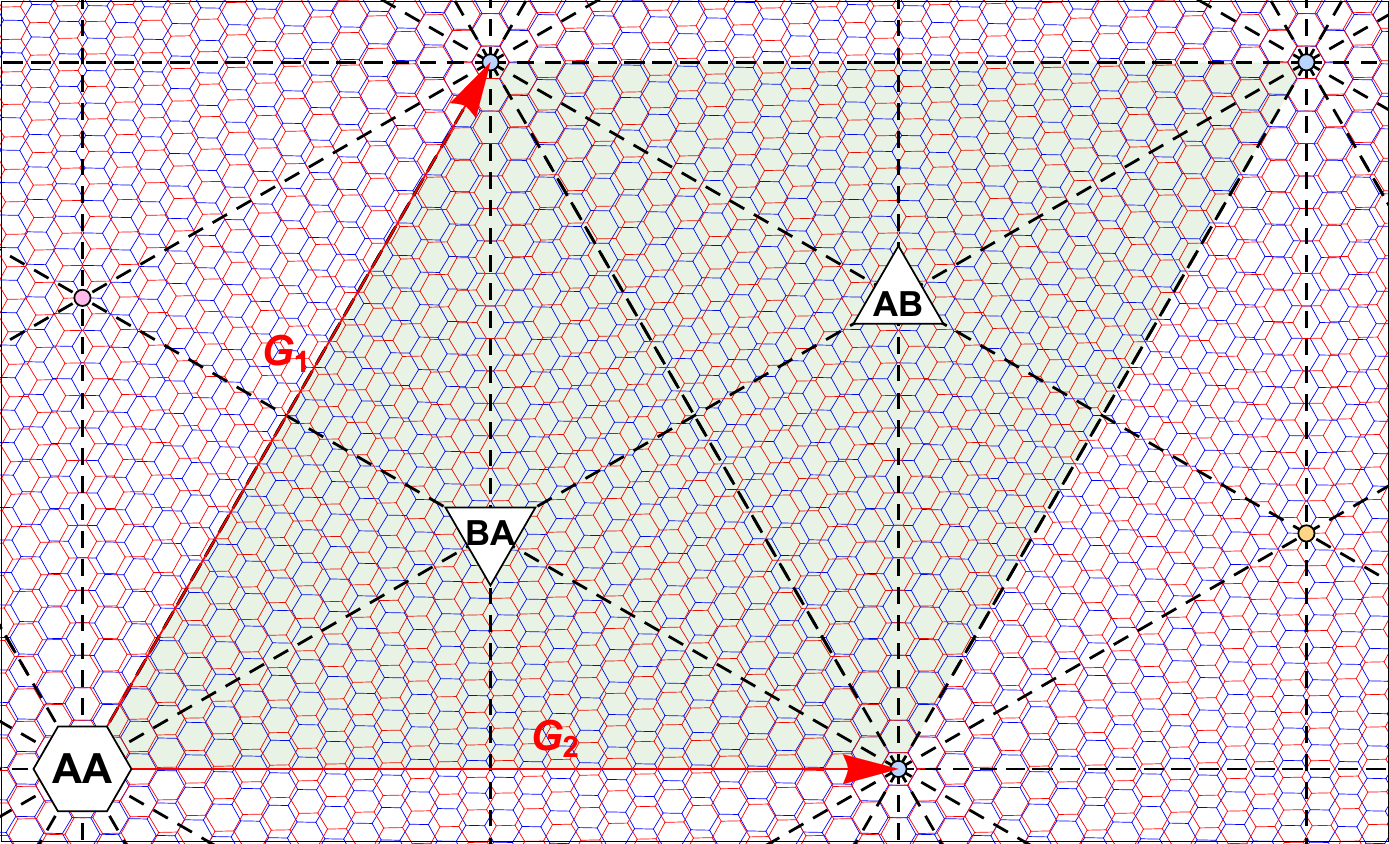}
	\caption{\label{FigLatticeSymmetries}
	The enlarged unit cell of twisted bilayer graphene (TBG), with the two different layers shown in red and blue. The new unit cell, spanned by vectors ${\bf G}_i$, has one six-fold rotation center with AA interlayer stacking, and two three-fold rotation centers with AB or BA stacking. There are {\em no} mirror symmetries: a reflection along one of the dashed lines interchanges the two layers. The corresponding space group of TBG is therefore the double cover of $p6$, as opposed to $p6m$ for single layer graphene or bilayers without twist.
	}
\end{figure}

\section{The model}\label{model}
Our aim is to understand the essential properties of twisted bilayer graphene band-structure close to charge neutrality. A single layer of graphene has a honeycomb structure, with lattice unit vectors ${\bf a}_{\pm} = \tfrac{a}{2} \left( \pm \hat{\bf x} + \sqrt{3} \hat{\bf y} \right) $ where $a = 0.246$ nm\cite{Reich:2002kd,CastroNeto:2009cl}. The unit cell contains two inequivalent sites, labeled A and B. The corresponding band structure is known for having Dirac cones at ${\bf K}$ and ${\bf K'}$. The symmetry group is $p6m$, which means the unit cell contains one 6-fold rotation center, two 3-fold rotation centers and six distinct reflection axes.

The band structure of a bilayer depends on how the two layers are stacked. A literal stacking of two layers on top of each other, where the A sites in one layer are on top of the A sites of the other, is known as AA stacking. A more natural stacking occurs when the A sites of one layer are above the B sites of the other layer, this is known as AB stacking. 

We make `twisted' bilayer graphene by starting with AB stacking and rotating one of the layers around an AB site. To achieve a commensurate rotation, we can choose two integers $m,n$ and define the rotation angle $\theta$ as the angle between $n{\bf a}_1 + m {\bf a}_2$ and $m {\bf a}_1 + n {\bf a}_2$\cite{SuarezMorell:2010bz}. The new unit cell has unit vectors ${\bf G}_1 = n {\bf a}_1 + m {\bf a}_2$ and ${\bf G}_2 = -m {\bf a}_1 + (n+m) {\bf a}_2$ and contains $4 (n^2 + nm + m^2)$ atoms. This new unit cell, shown in Fig. \ref{FigLatticeSymmetries}, possesses a region where the layers are effectively AA stacked, as well as regions with AB stacking and BA stacking - the latter just being the same as AB stacking but with the two layers exchanged. Note that the unit cell is {\em different} when one starts with AA stacking and rotates around one of the carbon atoms. In this case, the symmetry is actually lower when rotating with a small angle, because now the unit cell does not contain a sixfold rotation center.\footnote{We thank J.~W.~F.~Venderbos for pointing this out to us.}

Interestingly, the space group of the twisted bilayer is {\em different} from the single layer one. It does contain 6-fold rotations around the AA centers, and 3-fold rotations around the AB and BA centers. However, there are no mirror symmetries, as can be inferred from Fig. \ref{FigLatticeSymmetries}. Reflection along one of the symmetry axes interchanges the two layers. Therefore, the space group is the double cover of $p6$, rather than $p6m$.\cite{Dresselhaus:2007wd,Bradlyn:2017fy,Mele:2011cb}

\begin{figure}
	\includegraphics[width=\columnwidth]{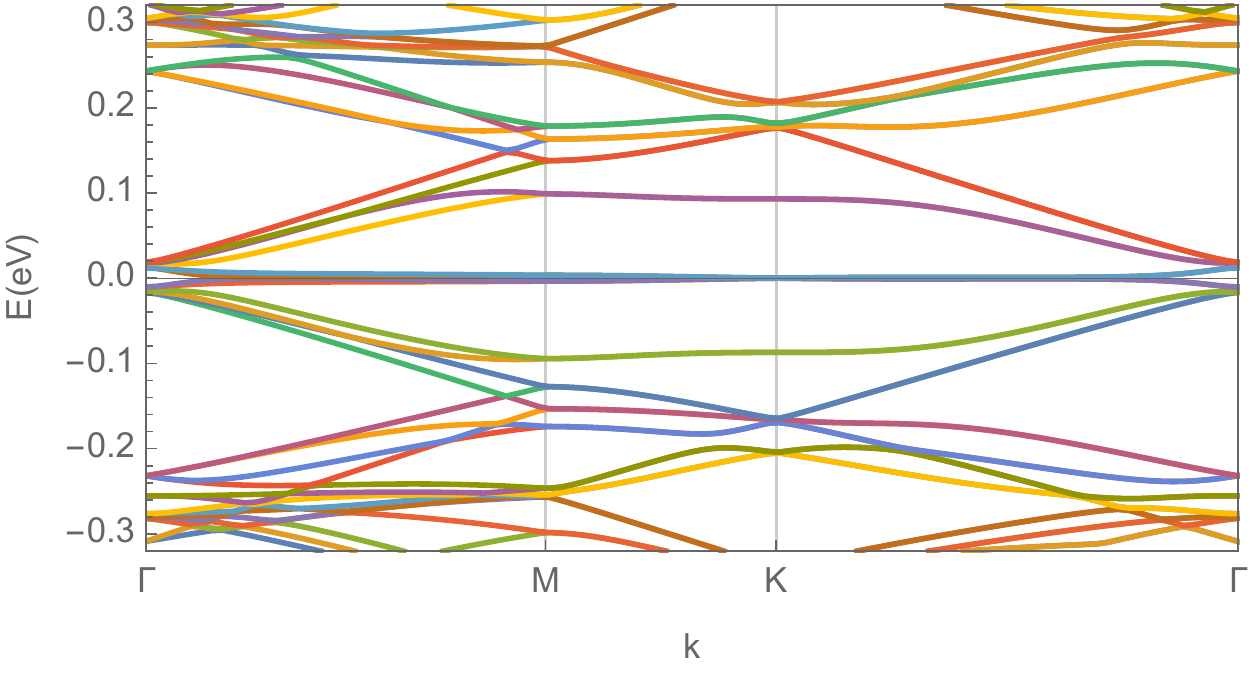}
	\includegraphics[width=\columnwidth]{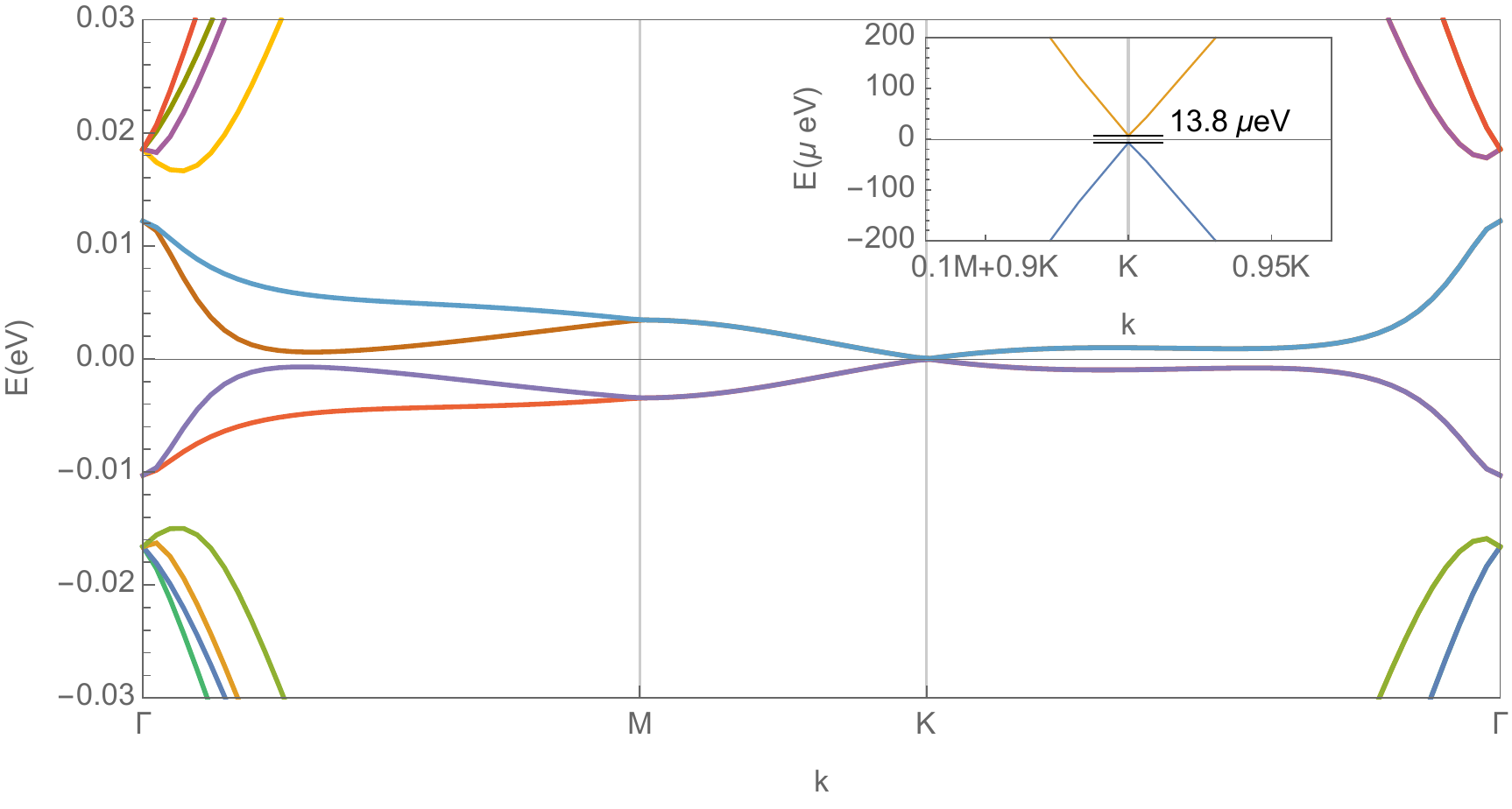}
	\caption{\label{FigFlatBands}		
	A tight-binding band structure for bilayer graphene with twist angle $\theta=1.08^\circ$, using the parameters provided in the main text. This model contains 11,164 bands. Four flat bands with a bandwidth $W = 11.25$ meV arise around charge neutrality, as is clearly shown in the bottom panel. At the ${\bf K}$-point there is a small gap of $13.8 \mu$eV at the approximate Dirac cones, as shown in the inset.}
\end{figure}

The magic angle of Refs. [\onlinecite{Cao:2018kn,Cao:2018ff}] was $\theta = 1.08$, which can be generated by choosing $m=31$ and $n=30$. In this case, the new unit cell contains 11,164 atoms. Needless to say, an exact description of the band structure from first principles is a very difficult task.\cite{Song:2018ul} However, we can write down a tight-binding model including all the 11,164 bands using parameters from literature. We choose the in-plane nearest neighbor hopping $t = 2.8$ eV and interlayer hopping described by $t_\perp({\bf r}) = t_{\perp0} e^{ - |{\bf r}|/\xi}$, where ${\bf r}$ is the total distance between two atoms including the interlayer distance $d = 0.335$ nm, $\xi = 0.11a$ and $t_{\perp0}$ is chosen such that $t_\perp = 0.35$ eV for the AA stacked atoms\cite{SuarezMorell:2010bz,Fang:2016iq,Jung:2014hj,Malard:2007jv}. The resulting band structure for the magic angle system is shown in Fig. \ref{FigFlatBands}. 

Continuum models predict double degenerate Dirac cones at the ${\bf K}$ and ${\bf K'}$ points, with drastically reduced velocity.\cite{Bistritzer:2011ho,LopesdosSantos:2012vk} However, a careful analysis shows that there is actually {\em not} a four-fold degeneracy of states at ${\bf K}$ and ${\bf K'}$. Instead, much like AA stacked bilayers, we find two double degenerate states separated by a minuscule gap of 13.8 $\mu$eV, as shown in Fig. \ref{FigFlatBands}. Note that the alternate way of creating twisted bilayers, by starting with AA stacking and rotating around one of the carbon atoms, yields a different degeneracy structure. In that case there is a double degenerate state at charge neutrality at ${\bf K}$, separated by a $\mu$eV gap to two single degenerate states above and below. This is the same degeneracy structure as in AB stacked bilayers. In both cases there are therefore only `approximate' Dirac cones, with two flat bands above and two flat bands below them. The bandwidth of these bands is severely reduced to about $W=11.25$ meV. Any effective model should include these correct degeneracies and the corresponding narrow bands. 

\begin{figure}
	\includegraphics[width=\columnwidth]{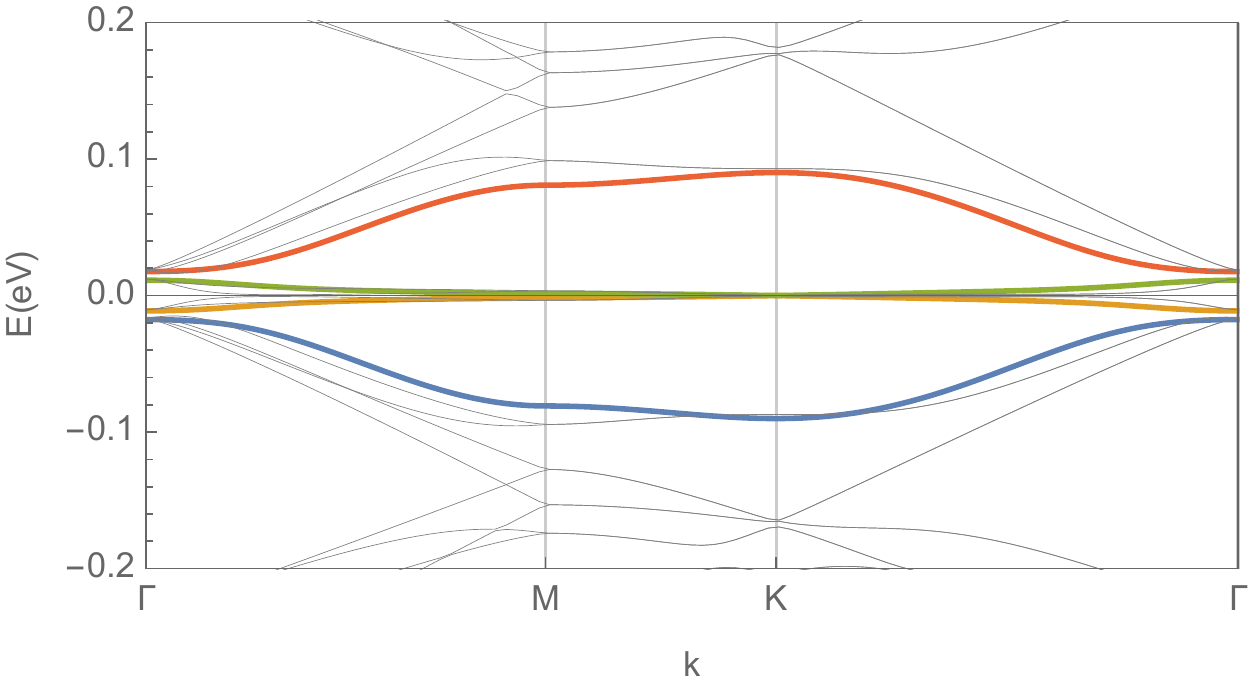}
	\caption{\label{FigEffBands}
	An effective 8-band model, where each band is double degenerate, describing the essential low-energy physics of the twisted bilayer, Eq. (5). 
	The fitted parameters are $t_K = 1.53$ meV, $t_\Gamma = 9.50$ meV, $\Delta_\Gamma = 61.60$ meV, and $t' = 4.67$ meV. The full band structure of Fig. \ref{FigFlatBands} is shown in grey.
	}
\end{figure}

To understand the Mott physics, however, a precise knowledge of the {\em real space structure} of the flat bands is necessary. We display the spatial structure of the low energy wavefunctions at ${\bf K}$ and ${\bf \Gamma}$ in Fig. \ref{FigOrbitals}. Consistent with tunnelling experiments,\cite{Wong:2015dg} the electrons close to charge neutrality are strongly localized at the AA regions. However, at the ${\bf \Gamma}$ point a clear transfer of charge is seen to a ring around the AA center. Any effective model of the whole band should capture this charge-transfer. The standard maximally localized Wannier orbitals, following the algorithm of Ref. [\onlinecite{Marzari:2012eu,Marzari:1997wa}], will be extended over more than one unit cell\cite{Po:2018,Zou:2018ud,Po:2018vk,Kang:2018wd}. To obtain the proper `ring' structure properly, one needs to occupy all orbitals in a unit cell and its six neighbors. In such a program, setting the correct interaction, requires six-orbital-interaction terms,\cite{Koshino:2018td} which unnecessarily complicates things.  Instead, by including more bands in the effective hamiltonian we can retrieve orbitals that are localized {\em within one unit cell}.\cite{Po:2018vk,Song:2018ul}

Let us focus first on the ${\bf K}$-orbitals or `center' orbitals, whose weight at the AA centers form a triangular lattice. The ${\bf K}$ and ${\bf K'}$ are exchanged under six- and two-fold rotations with respect to the $z$ axis perpendicular to the TBG plane. As a direct consequence, the little co-group at the ${\bf K}$-point is $C_3$, a subgroup of $C_6$, the point group of TBG. Orbitals at this point could realize irrep $A$ and irrep $E$ of $C_3$. 
Consider the equivalent representation at ${\bf K}$ which involves the transformation of $\exp(i {\bf K} {\bf R}_j)$ under the symmetry operations of $C_3$. Rotation of the radius vector ${\bf R}_j$ by the angle $2\pi/3$ anticlockwise is equivalent to rotation of the vector ${\bf K}$ in the opposite direction, that is to substitution of the three equivalent corners of the small Brillouin zone. 
When this reducible representation is decomposed into irreps of the group $C_3$, we find that it is exactly contained in the two-dimensional irrep $E$ of $C_3$. Therefore orbitals at the ${\bf K}$ point should be two-fold degenerate. 
However, because the gap at the approximate Dirac cones is practically unobservable, the effective hopping among ${\bf K}$ (center) orbitals can be approximated by a honeycomb symmetry as is done in \cite{Zhang:2018,Yuan:2018un,Po:2018,Zou:2018ud,Po:2018vk,Kang:2018wd}. The ${\bf \Gamma}$-orbitals, on the other hand, are far away from the approximate Dirac cones and will be treated as having hopping on a triangular lattice.

Instead of using the Wannier construction to build an exact copy of the tight binding bandstructure, we will introduce a model that qualitatively captures the low-energy physics. Note that the precise tight-binding band structure depends on specific model details, including a possible lattice relaxation.\cite{Choi:2018vw,Angeli:2018wy} We therefore refrain from claiming to be numerically exact and instead focus on the relevant physical mechanisms. In the absence of hybridization between `ring' and `center' orbitals, we have four gapped triangular-lattice orbitals at the ${\bf \Gamma}$ point and four-fold degenerate Dirac cones. Hybridization between ${\bf \Gamma}$ and ${\bf K}$-orbitals causes the lowest energy band to be a mix of both orbitals. To construct an effective model we split the 8 lowest energy bands in two degenerate `valleys'. Each `valley' consists of two ring and two center-orbitals, and there are four parameters that determine the band structure: hopping between same type of orbitals $t_K$ and $t_\Gamma$, hybridization between different type orbitals $t'$, and the gap of the ${\bf \Gamma}$-orbitals $\Delta_\Gamma$. The resulting band-structure is shown in Fig. \ref{FigEffBands}, and this model will serve as the starting point for our analysis of the insulating phase.

Explicitly, the Hamiltonian of the effective model consists of two degenerate $4 \times 4$-blocks. The honeycomb symmetry of the ${\bf K}$-orbitals is reflected in the hopping factors
\begin{equation}
	f_K (\bk) = 1 + e^{i {\bf a}_1 \cdot \bk} + e^{i {\bf a}_2 \cdot \bk}
\end{equation}
whereas the triangular nature of the ${\bf \Gamma}$-orbitals is realized by the factors
\begin{equation}
	f_\Gamma(\bk) = 2 \left( \cos {\bf a}_1 \cdot \bk + \cos {\bf a}_2 \cdot \bk + \cos {\bf a}_3 \cdot \bk  \right).
\end{equation}
The coupling between ${\bf K}$ and ${\bf \Gamma}$-orbitals can can take two forms,
\begin{eqnarray}
	f_{K\Gamma_1} (\bk) & = & 1 + e^{- i {\bf a}_1 \cdot \bk} + e^{- i {\bf a}_2 \cdot \bk}, \\
	f_{K\Gamma_2} (\bk) & = & e^{-i ( {\bf a}_1 + {\bf a}_2 ) \cdot \bk} + e^{- i {\bf a}_1 \cdot \bk} + e^{- i {\bf a}_2 \cdot \bk},
\end{eqnarray}
reflecting the two possible ways to combine honeycomb and triangular symmetries. Thus, each $4 \times 4$-block of the effective Hamiltonian reads
\begin{widetext}
\begin{equation}
	H_4 (\bk) = 
		\begin{pmatrix}
			0 						& t_K f_K (\bk) 	& t' f_{K\Gamma_1} (\bk)	& 0						 \\
			t_K f_K (-\bk)		& 0					 	& 0						& - t' f_{K\Gamma_2}(\bk) 	\\
			t' f_{K\Gamma_1} (-\bk) 	& 0					 	& -\Delta_\Gamma + t_\Gamma f_{\Gamma} (\bk)		& 0 \\
			0					 	& - t' f_{K\Gamma_2}(-\bk)	& 0						& \Delta_\Gamma - t_\Gamma f_{\Gamma} (\bk)
		\end{pmatrix}
\end{equation}
\end{widetext}
Using Eqn.(5) we fit the low energy bands of the full band structure by using the following parameters,
\begin{eqnarray}
	t_K 	& = & 1.5312 \, \mathrm{meV}, \\
	t_\Gamma 	& = & 9.5007 \, \mathrm{meV}, \\
	t' 		& = & 4.6730\, \mathrm{meV}, \\
	\Delta_\Gamma & = & 61.5978 \, \mathrm{meV}.
\end{eqnarray}
With these parameters we have an effective low-energy model with {\em fully localized} orbitals, that describe the four flat bands as well as four dispersive bands that are further away from charge-neutrality.

\begin{figure}
	\includegraphics[width=0.5\columnwidth]{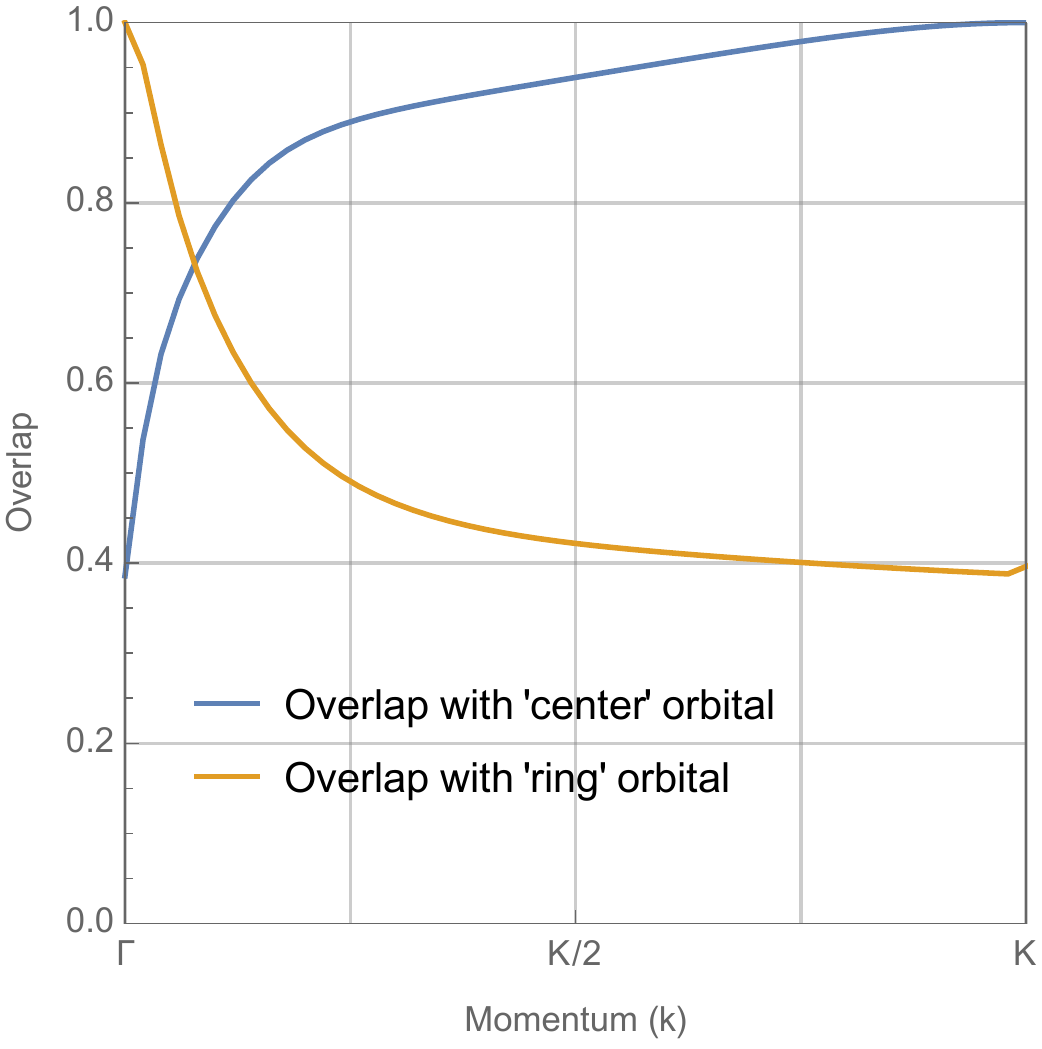}
	\caption{\label{FigOverlap}
	Overlap between the eigenstates of the flat band from the full band problem, with ring orbitals in yellow, and center orbitals in blue. Note that any filling away from charge-neutrality leads to an uneven charge distribution in the unit cell, as is shown in Fig. \ref{FigChargeTransfer}.
}
\end{figure}

Note that we have discussed the orbital character of the low-energy states at exactly ${\bf \Gamma}$ and ${\bf K}$. At momenta in between these two high energy points, the orbital character smoothly transforms from ring-like to center-like, as is shown in Fig. \ref{FigOverlap}. The overlap is defined as the norm of the wavefunction at momentum $\bk$ projected onto the subspace of the four lowest energy wavefunctions at either ${\bf \Gamma}$ (`ring') or ${\bf K}$ (`center'). Note that the overlap between ring and center orbitals is $\approx 0.4$. 

\section{Interactions and the Mott phase} \label{interactions} 

The emergence of insulating behavior, when non-interacting theories predict conducting, can be due to either Wigner or Mott localization. A Wigner crystal\cite{Padhi:2018} can be dismissed due to commensurability\cite{Cao:2018kn}. Wigner crystals can exist in lattice systems at sufficiently low densities, but its wavelength changes with changing charge density, see for example Ref. [\onlinecite{Rademaker:2013jg}] for square lattice results. However, this is not the case in TBG: the insulating phase {\em only} appears at half-filling of the flat bands. Consequently, the insulating phase has a Mott character. 

A Mott insulator can be realized by adding to the tight-binding model an onsite Hubbard repulsion for each localized orbital, and it is known that single-layer graphene has a relatively strong onsite interaction $U = 9.3$ eV\cite{Schuler:2013ip,Wehling:2011cf}. However, for orbitals that span thousands of different atoms, the full Coulomb interaction beyond the onsite repulsion plays a central role. Indeed, the Coulomb energy is nonzero whenever there are {\em macroscopic charge inhomogeneities} $\delta n(\r)$,
\begin{equation}
	E_{\mathrm{int}} = \sum_{\r, \r'} \delta n(\r) \frac{e^2}{4 \pi \epsilon(\r-\r') |\r - \r'|} \delta n(\r')
	\label{IntEnergy}
\end{equation}
where $\delta n(\r)$ measures the deviation from the average electron density $\overline{n}$ at position $\r$. At charge neutrality, the electron charge density is evenly distributed over all carbon atoms. Because the charge density away from charge-neutrality is unevenly distributed, there will be a large `classical' contribution to the Coulomb energy proportional to $\langle \delta n(\r) \rangle V(\r - \r') \langle \delta n(\r') \rangle$. On the other hand, the usual `quantum' Hubbard contribution proportional $\langle  n_{i\uparrow}  n_{i\downarrow} \rangle$ will be smaller as the charge fluctuations on the unit cell scale will be small, of the order of $1/\sqrt{A} \approx \theta$, where $A$ is the area of the unit cell. 

\begin{figure}
	\includegraphics[width=0.49\columnwidth]{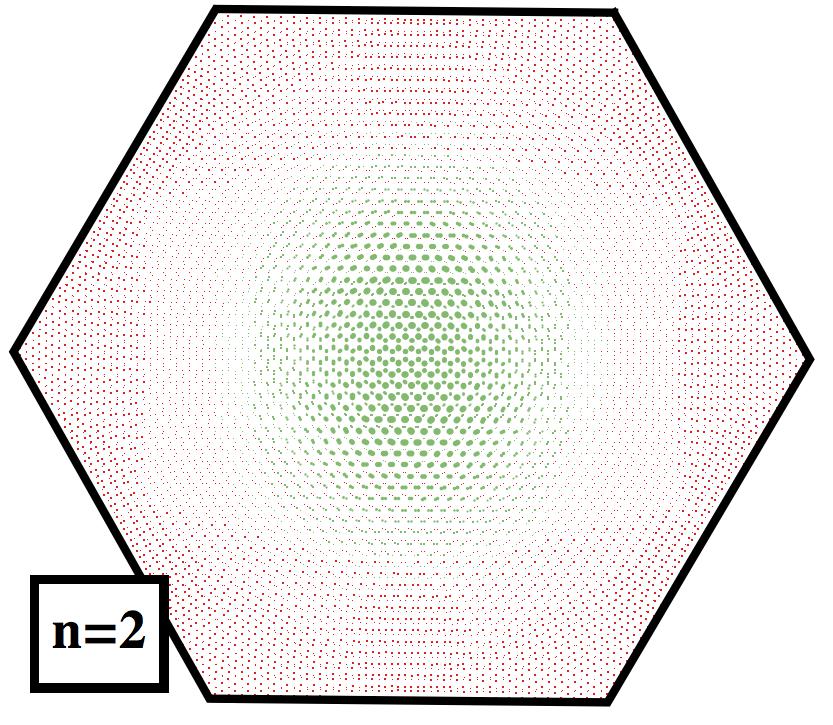}
	\includegraphics[width=0.49\columnwidth]{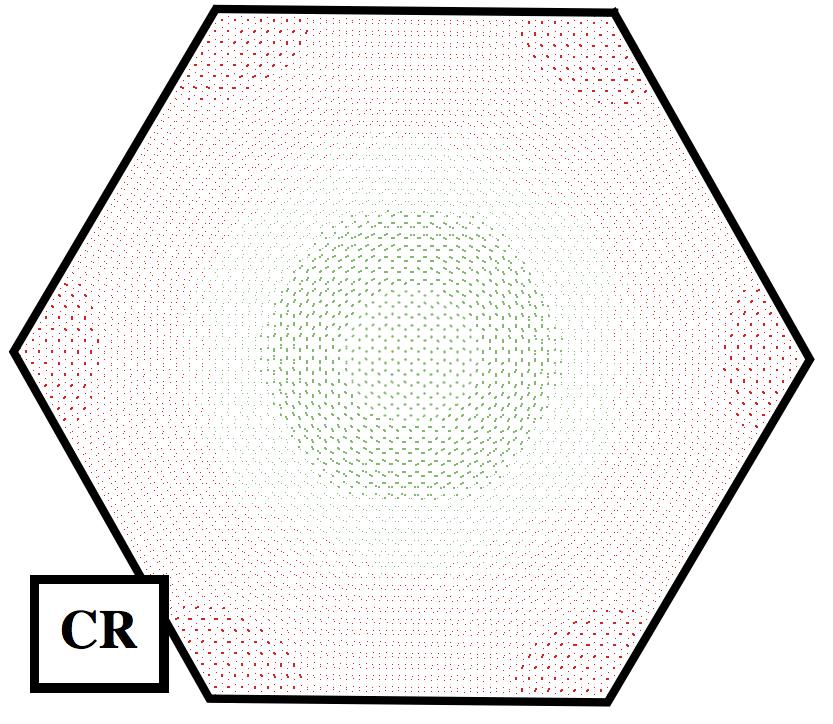}
	\includegraphics[width=\columnwidth]{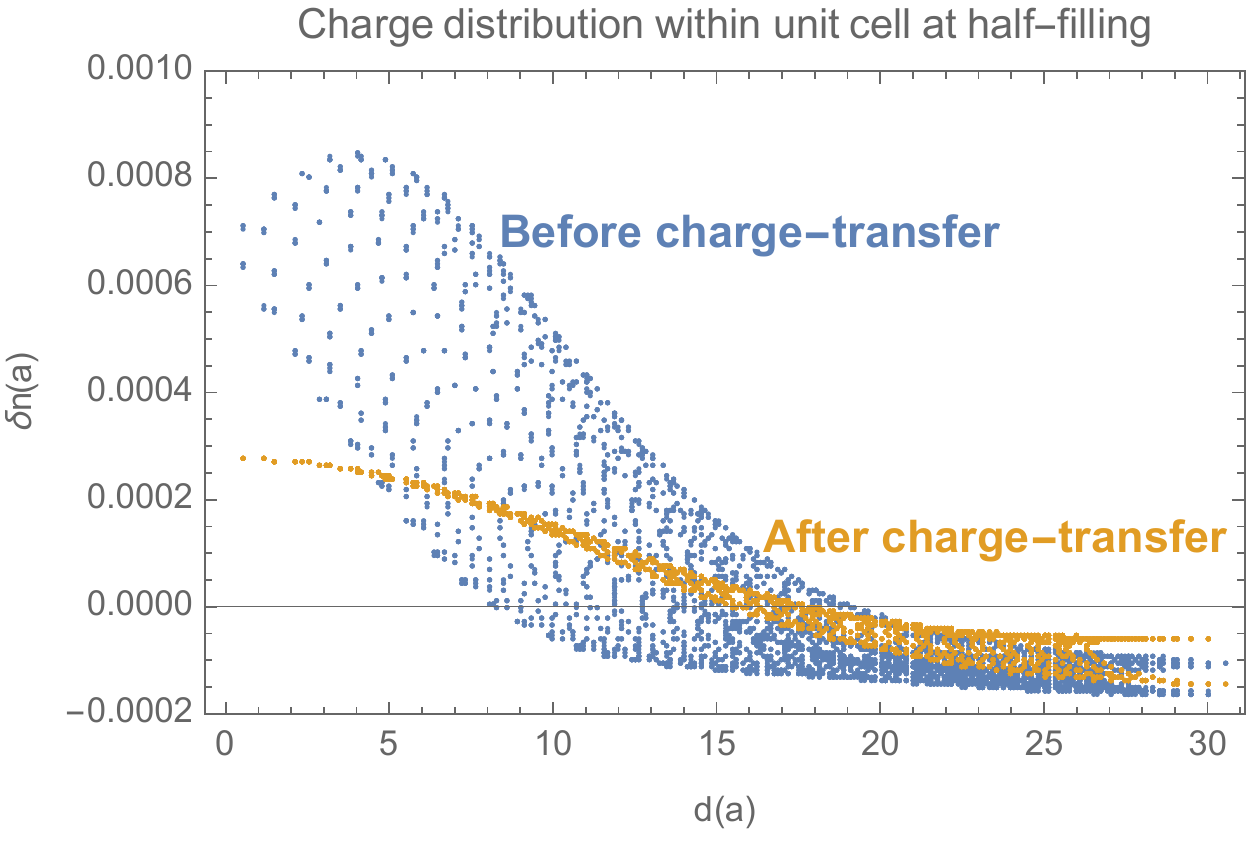}
	\caption{\label{FigChargeTransfer}
	A comparison of the charge distribution within one unit cell in the non-interacting model at half-filling of the flat band (top left) versus the charge-transfer scenario (top right). Shown here is in green an excess of charge and in red an absence of charge. The charge density per carbon atom, as a function of distance from the unit cell center, is shown in the bottom panel. In the non-interacting case the Coulomb energy is $E_{\mathrm{int}} = 317.6$ meV, whereas in the charge-transfer picture this is reduced to $E_{\mathrm{int}} = 103.9$ meV.
}
\end{figure}

As experimentally observed,\cite{Wong:2015dg} upon doping away from charge neutrality the electronic charge density will cluster around the AA centers. We visualize this in the top left panel of Fig. \ref{FigChargeTransfer}, where the charge distribution in one unit cell is shown at a density of two holes relative to charge neutrality using the full band structure of Fig. \ref{FigFlatBands}. A large concentration of green dots (excess charge) is seen in the AA centers, whereas around the AB/BA centers there is a depletion of charge (red dots). Indeed, we found that about 91\% of the added charge ends up in the `center'-like ${\bf K}$-orbitals.

According to Ref. [\onlinecite{Wehling:2011cf}], interactions in single layer graphene are strong:  the effective on-site (Hubbard) interaction was found to be $U = 9.3$ eV  while the nearest neighbor Coulomb interaction strength was computed to $V = 5.5$ eV. The long-wavelength limit of the dielectric constant approaches $\epsilon(k=0) = 1$ and therefore we expect to find genuine long-range interactions $(1/r)$. The form suggested by Wehling et al. prompts us to use the following effective form of the screened Coulomb interaction in the tight-binding model,
\begin{equation}
	V(\r_i-\r_j) = \frac{1.438}{ 0.116 + |\r_i - \r_j|} \; \mathrm{eV}
\end{equation}
where the distance between two carbon atoms $|\r_i - \r_j|$ should be measured in nm. The resulting classical Coulomb energy at half-filling of the flat band is enormous: $E_{\mathrm{int}} = 317.6$ meV! 

Much of this Coulomb energy can be reduced by a {\em charge-transfer} from the `center' (${\bf K}$) to the `ring' (${\bf \Gamma}$) orbitals. A zeroeth order model of such a charge-transfer scenario would have exactly one localized hole in the center orbital and one localized hole in the ring orbital. Compared to the non-interacting case we lose the kinetic energy, estimated at 7.86 meV, and we need to pay the gap $\Delta_{\Gamma}=61.6$ meV of the ring orbital. However, the interaction energy is drastically reduced to $E_{\mathrm{int}} = 103.9$ meV. Overall, the average energy gain per unit cell due to the charge transfer is estimated at $\Delta E = 154$ meV. 

This is the core prediction of our paper: when doping away from charge neutrality a charge-transfer occurs from the center to the ring orbitals because Coulomb repulsion wants to smoothen out charge. Notice that this is very similar to the physics of cuprates, where a charge-transfer from the copper $d$ to the oxygen $p$ orbitals causes insulating behavior\cite{Zaanen:1985bsa}.

Having settled one localized charge in the ${\bf K}$-orbital and one in the ${\bf \Gamma}$-orbital, the spin degree of freedom remains to be determined. In multi-orbital atomic Mott insulators the Hund's coupling favors a large total spin in the unit cell\cite{Imada:1998er,Georges:2013ju}. However, in TBG there is a nonzero  overlap between `ring' and `center' orbitals, which in our effective model yielded $t' = 4.67$ meV. After localizing one charge in a ring and one in a center orbital, we get an effective {\em antiferromagnetic} Heisenberg coupling between their spins in 2nd order perturbation theory
\begin{equation}
	H_{\mathrm{eff}} = J \sum_{i} \vec{S}_{i {\bf K}} \cdot \vec{S}_{i {\bf \Gamma}}
\end{equation}
where the effective exchange constant $J$ is given by
\begin{equation}
	J = 2 \frac{|t'|^2}{\Delta E} =  0.28 \mathrm{\, meV}.
\end{equation}
Contrary to Hund's expectations, the two spins in each unit cell will therefore form a {\em singlet}, which is the natural ground state for a model with an even number of spins per unit cell\cite{Hastings:2004cd}. We conclude that the Mott phase in TBG is a non-entangled featureless spin-singlet paramagnet, consistent with recent experimental measurements\cite{Cao:2018kn}. Note that that the value of $J$ is 3.3 Kelvin, consistent with the energy scale required to break the insulating state with either thermal excitations or an external magnetic field.\cite{Cao:2018ff,Cao:2018kn,Yankowitz:2018tx}

Similarly, second order perturbation theory suggests a ferromagnetic Heisenberg-like coupling for the orbital degrees of freedom.




\section{Superconductivity} \label{superconductivity}

Consider the insulating state that has two electrons per unit cell removed relative to charge neutrality. When doping away from this Mott state, the dynamics of the dopants is described by a $t-J$ model. A key feature of such models is the effective {\em nearest-neighbor attraction} between dopants.\cite{Spaiek:1988cz,1973PhRvB...8.2236K} This attraction can lead to superconductivity of the dopants. The symmetry of the pairing state depends crucially on the nature of the dopants themselves. 

A prominent consequence of the proposed charge-transfer is the difference between electron and hole-doping relative to the insulating phase. Adding electrons to the insulating state - that is, moving closer to charge neutrality - will add dopants on the ${\bf \Gamma}$-orbitals. The effective model will consist of a nearest-neighbor attraction on a triangular lattice. This has been studied before and the most likely superconducting state would be spin-singlet $d+id$-wave\cite{Chen:2013hg,Guo:2018}.

On the other hand, hole-doping adds carriers to the ${\bf K}$-orbitals, which realize an effective honeycomb lattice. Nearest-neighbor attraction on a honeycomb lattice leads to  exotic spin singlet $p+ip$-wave superconductivity, as was proposed for single layer graphene away from charge neutrality\cite{Uchoa:2007bq}. A symmetry difference between the electron and hole-doped superconducting phases relative to the Mott state would be a clear proof of the charge-transfer occurring in TBG.




\section{Outlook}

We showed that the observed insulating state in TBG can be described in terms of charge-transfer from `center' to `ring' orbitals around the region of AA stacking. An experimental signature of this transfer can be found in tunnelling experiments: the electron density at the AA region center should be less than expected based on non-interacting theories. Note that the lattice relaxation of the Moir\'{e} patterns due to electron-phonon coupling can influence the expected charge density\cite{Slotman:2015kd,vanWijk:2015bc,Nam:2017jh,Choi:2018vw,Angeli:2018wy}. Therefore a full first-principles computation of the interactions in magic angle twisted bilayers is necessary to quantify the suggested charge-transfer.

A direct consequence of the charge-transfer is that the Mott phase is a featureless spin singlet paramagnet. The lowest energy spin excitations will be propagating triplets, which could be observed using thin film resonant inelastic X-ray scattering\cite{Ament:2011jy}. Also the symmetry difference between electron- and hole-doped superconductors is a result of the charge-transfer, and should be observable in experiments similar to the phase-sensitive experiments in cuprates.\cite{VanHarlingen:1995ht}

Finally, we want to emphasize that we based our predictions on a simple analysis of the real space wavefunctions of an 11,164-bands model. The effective model is a hybrid mixture of triangular and honeycomb symmetries, and it is not a trivial task to construct a low energy effective $t-J$ model out of those ingredients. However, we think that developing such a model and studying it using both analytical and numerical methods might provide key insights towards the understanding of twisted bilayer graphene.

{\em Note - } After completion of this manuscript we became aware of two other papers that argued for the inclusion of more than just 4 bands in an effective low-energy description of twisted bilayer graphene.\cite{Song:2018ul,Po:2018vk}

\acknowledgments \emph{Acknowledgments} - We are thankful to T.~Hsieh, J.~W.~F.~Venderbos, M.~I.~Katsnelson, P.~W.~Phillips, O.~Vafek, W.~Ku and V.~Dobrosavljevi\'{c} for discussions.
P.~M. acknowledges Fondecyt Grant No. 1160239. 
L.~R. is supported by the SNSF by an Ambizione grant.
This research was supported in part by Perimeter Institute for Theoretical Physics. Research at Perimeter Institute is supported by the Government of Canada through the Department of Innovation, Science, and Economic Development, and by the Province of Ontario through the Ministry of Research and Innovation.


\begin{thebibliography}{71}
\expandafter\ifx\csname natexlab\endcsname\relax\def\natexlab#1{#1}\fi
\expandafter\ifx\csname bibnamefont\endcsname\relax
  \def\bibnamefont#1{#1}\fi
\expandafter\ifx\csname bibfnamefont\endcsname\relax
  \def\bibfnamefont#1{#1}\fi
\expandafter\ifx\csname citenamefont\endcsname\relax
  \def\citenamefont#1{#1}\fi
\expandafter\ifx\csname url\endcsname\relax
  \def\url#1{\texttt{#1}}\fi
\expandafter\ifx\csname urlprefix\endcsname\relax\def\urlprefix{URL }\fi
\providecommand{\bibinfo}[2]{#2}
\providecommand{\eprint}[2][]{\url{#2}}

\bibitem[{\citenamefont{Cao et~al.}(2018{\natexlab{a}})\citenamefont{Cao,
  Fatemi, Demir, Fang, Tomarken, Luo, Sanchez-Yamagishi, Watanabe, Taniguchi,
  Kaxiras et~al.}}]{Cao:2018kn}
\bibinfo{author}{\bibfnamefont{Y.}~\bibnamefont{Cao}},
  \bibinfo{author}{\bibfnamefont{V.}~\bibnamefont{Fatemi}},
  \bibinfo{author}{\bibfnamefont{A.}~\bibnamefont{Demir}},
  \bibinfo{author}{\bibfnamefont{S.}~\bibnamefont{Fang}},
  \bibinfo{author}{\bibfnamefont{S.~L.} \bibnamefont{Tomarken}},
  \bibinfo{author}{\bibfnamefont{J.~Y.} \bibnamefont{Luo}},
  \bibinfo{author}{\bibfnamefont{J.~D.} \bibnamefont{Sanchez-Yamagishi}},
  \bibinfo{author}{\bibfnamefont{K.}~\bibnamefont{Watanabe}},
  \bibinfo{author}{\bibfnamefont{T.}~\bibnamefont{Taniguchi}},
  \bibinfo{author}{\bibfnamefont{E.}~\bibnamefont{Kaxiras}},
  \bibnamefont{et~al.}, \bibinfo{journal}{Nature}
  \textbf{\bibinfo{volume}{556}}, \bibinfo{pages}{80}
  (\bibinfo{year}{2018}{\natexlab{a}}).

\bibitem[{\citenamefont{Cao et~al.}(2018{\natexlab{b}})\citenamefont{Cao,
  Fatemi, Fang, Watanabe, Taniguchi, Kaxiras, and
  Jarillo-Herrero}}]{Cao:2018ff}
\bibinfo{author}{\bibfnamefont{Y.}~\bibnamefont{Cao}},
  \bibinfo{author}{\bibfnamefont{V.}~\bibnamefont{Fatemi}},
  \bibinfo{author}{\bibfnamefont{S.}~\bibnamefont{Fang}},
  \bibinfo{author}{\bibfnamefont{K.}~\bibnamefont{Watanabe}},
  \bibinfo{author}{\bibfnamefont{T.}~\bibnamefont{Taniguchi}},
  \bibinfo{author}{\bibfnamefont{E.}~\bibnamefont{Kaxiras}}, \bibnamefont{and}
  \bibinfo{author}{\bibfnamefont{P.}~\bibnamefont{Jarillo-Herrero}},
  \bibinfo{journal}{Nature} \textbf{\bibinfo{volume}{556}}, \bibinfo{pages}{43}
  (\bibinfo{year}{2018}{\natexlab{b}}).

\bibitem[{\citenamefont{Yankowitz et~al.}(2018)\citenamefont{Yankowitz, Chen,
  Polshyn, Watanabe, Taniguchi, Graf, Young, and Dean}}]{Yankowitz:2018tx}
\bibinfo{author}{\bibfnamefont{M.}~\bibnamefont{Yankowitz}},
  \bibinfo{author}{\bibfnamefont{S.}~\bibnamefont{Chen}},
  \bibinfo{author}{\bibfnamefont{H.}~\bibnamefont{Polshyn}},
  \bibinfo{author}{\bibfnamefont{K.}~\bibnamefont{Watanabe}},
  \bibinfo{author}{\bibfnamefont{T.}~\bibnamefont{Taniguchi}},
  \bibinfo{author}{\bibfnamefont{D.}~\bibnamefont{Graf}},
  \bibinfo{author}{\bibfnamefont{A.~F.} \bibnamefont{Young}}, \bibnamefont{and}
  \bibinfo{author}{\bibfnamefont{C.~R.} \bibnamefont{Dean}},
  \bibinfo{journal}{arXiv}  (\bibinfo{year}{2018}), \eprint{1808.07865v2}.

\bibitem[{\citenamefont{Efimkin and MacDonald}(2018)}]{Efimkin:2018us}
\bibinfo{author}{\bibfnamefont{D.~K.} \bibnamefont{Efimkin}} \bibnamefont{and}
  \bibinfo{author}{\bibfnamefont{A.~H.} \bibnamefont{MacDonald}},
  \bibinfo{journal}{arXiv}  (\bibinfo{year}{2018}), \eprint{1803.06404v1}.

\bibitem[{\citenamefont{Pal et~al.}(2018)\citenamefont{Pal, Spitz, and
  Kindermann}}]{Pal:2018wc}
\bibinfo{author}{\bibfnamefont{H.~K.} \bibnamefont{Pal}},
  \bibinfo{author}{\bibfnamefont{S.}~\bibnamefont{Spitz}}, \bibnamefont{and}
  \bibinfo{author}{\bibfnamefont{M.}~\bibnamefont{Kindermann}},
  \bibinfo{journal}{arXiv}  (\bibinfo{year}{2018}), \eprint{1803.07060v1}.

\bibitem[{\citenamefont{Xu and Balents}(2018)}]{Xu:2018}
\bibinfo{author}{\bibfnamefont{C.}~\bibnamefont{Xu}} \bibnamefont{and}
  \bibinfo{author}{\bibfnamefont{L.}~\bibnamefont{Balents}},
  \bibinfo{journal}{arXiv}
  (\bibinfo{year}{2018}), \eprint{1803.08057v2}.

\bibitem[{\citenamefont{Yuan and Fu}(2018)}]{Yuan:2018un}
\bibinfo{author}{\bibfnamefont{N.~F.~Q.} \bibnamefont{Yuan}} \bibnamefont{and}
  \bibinfo{author}{\bibfnamefont{L.}~\bibnamefont{Fu}},
  \bibinfo{journal}{arXiv}  (\bibinfo{year}{2018}), \eprint{1803.09699v1}.

\bibitem[{\citenamefont{Po et~al.}(2018{\natexlab{a}})\citenamefont{Po, Zou,
  Vishwanath, and Senthil}}]{Po:2018}
\bibinfo{author}{\bibfnamefont{H.~C.} \bibnamefont{Po}},
  \bibinfo{author}{\bibfnamefont{L.}~\bibnamefont{Zou}},
  \bibinfo{author}{\bibfnamefont{A.}~\bibnamefont{Vishwanath}},
  \bibnamefont{and} \bibinfo{author}{\bibfnamefont{T.}~\bibnamefont{Senthil}},
  \bibinfo{journal}{arXiv}
  (\bibinfo{year}{2018}{\natexlab{a}}), \eprint{1803.09742v1}.

\bibitem[{\citenamefont{Roy and Juricic}(2018)}]{Roy:2018}
\bibinfo{author}{\bibfnamefont{B.}~\bibnamefont{Roy}} \bibnamefont{and}
  \bibinfo{author}{\bibfnamefont{V.}~\bibnamefont{Juricic}},
  \bibinfo{journal}{arXiv} 
  (\bibinfo{year}{2018}), \eprint{1803.11190v1}.

\bibitem[{\citenamefont{Guo et~al.}(2018)\citenamefont{Guo, Zhu, Feng, and
  Scalettar}}]{Guo:2018}
\bibinfo{author}{\bibfnamefont{H.}~\bibnamefont{Guo}},
  \bibinfo{author}{\bibfnamefont{X.}~\bibnamefont{Zhu}},
  \bibinfo{author}{\bibfnamefont{S.}~\bibnamefont{Feng}}, \bibnamefont{and}
  \bibinfo{author}{\bibfnamefont{R.~T.} \bibnamefont{Scalettar}},
  \bibinfo{journal}{arXiv} 
  (\bibinfo{year}{2018}), \eprint{1804.00159v1}.

\bibitem[{\citenamefont{Padhi et~al.}(2018)\citenamefont{Padhi, Setty, and
  Phillips}}]{Padhi:2018}
\bibinfo{author}{\bibfnamefont{B.}~\bibnamefont{Padhi}},
  \bibinfo{author}{\bibfnamefont{C.}~\bibnamefont{Setty}}, \bibnamefont{and}
  \bibinfo{author}{\bibfnamefont{P.~W.} \bibnamefont{Phillips}},
  \bibinfo{journal}{arXiv}
  (\bibinfo{year}{2018}), \eprint{1804.01101v1}.

\bibitem[{\citenamefont{Irkhin and Skryabin}(2018)}]{Irkhin:2018}
\bibinfo{author}{\bibfnamefont{V.~Y.} \bibnamefont{Irkhin}} \bibnamefont{and}
  \bibinfo{author}{\bibfnamefont{Y.~N.} \bibnamefont{Skryabin}},
  \bibinfo{journal}{arXiv} 
  (\bibinfo{year}{2018}), \eprint{1804.02236v2}.

\bibitem[{\citenamefont{Dodaro et~al.}(2018)\citenamefont{Dodaro, Kivelson,
  Schattner, Sun, and Wang}}]{Dodaro:2018}
\bibinfo{author}{\bibfnamefont{J.~F.} \bibnamefont{Dodaro}},
  \bibinfo{author}{\bibfnamefont{S.~A.} \bibnamefont{Kivelson}},
  \bibinfo{author}{\bibfnamefont{Y.}~\bibnamefont{Schattner}},
  \bibinfo{author}{\bibfnamefont{X.-Q.} \bibnamefont{Sun}}, \bibnamefont{and}
  \bibinfo{author}{\bibfnamefont{C.}~\bibnamefont{Wang}},
  \bibinfo{journal}{arXiv} 
  (\bibinfo{year}{2018}), \eprint{1804.03162v2}.

\bibitem[{\citenamefont{Zhang}(2018)}]{Zhang:2018}
\bibinfo{author}{\bibfnamefont{L.}~\bibnamefont{Zhang}},
  \bibinfo{journal}{arXiv} 
  (\bibinfo{year}{2018}), \eprint{1804.09047}.

\bibitem[{\citenamefont{Zhang et~al.}(2018)\citenamefont{Zhang, Mao, Cao,
  Jarillo-Herrero, and Senthil}}]{Zhang:2018wj}
\bibinfo{author}{\bibfnamefont{Y.-H.} \bibnamefont{Zhang}},
  \bibinfo{author}{\bibfnamefont{D.}~\bibnamefont{Mao}},
  \bibinfo{author}{\bibfnamefont{Y.}~\bibnamefont{Cao}},
  \bibinfo{author}{\bibfnamefont{P.}~\bibnamefont{Jarillo-Herrero}},
  \bibnamefont{and} \bibinfo{author}{\bibfnamefont{T.}~\bibnamefont{Senthil}},
  \bibinfo{journal}{arXiv}
  (\bibinfo{year}{2018}), \eprint{1805.08232}.

\bibitem[{\citenamefont{Thomson et~al.}(2018)\citenamefont{Thomson, Chatterjee,
  Sachdev, and Scheurer}}]{Thomson:2018ve}
\bibinfo{author}{\bibfnamefont{A.}~\bibnamefont{Thomson}},
  \bibinfo{author}{\bibfnamefont{S.}~\bibnamefont{Chatterjee}},
  \bibinfo{author}{\bibfnamefont{S.}~\bibnamefont{Sachdev}}, \bibnamefont{and}
  \bibinfo{author}{\bibfnamefont{M.~S.} \bibnamefont{Scheurer}},
  \bibinfo{journal}{arXiv}
  (\bibinfo{year}{2018}), \eprint{1806.02837}.

\bibitem[{\citenamefont{Su and Lin}(2018)}]{Su:2018vz}
\bibinfo{author}{\bibfnamefont{Y.}~\bibnamefont{Su}} \bibnamefont{and}
  \bibinfo{author}{\bibfnamefont{S.-Z.} \bibnamefont{Lin}},
  \bibinfo{journal}{arXiv}
  (\bibinfo{year}{2018}), \eprint{1807.02196}.

\bibitem[{\citenamefont{Zou et~al.}(2018)\citenamefont{Zou, Po, Vishwanath, and
  Senthil}}]{Zou:2018ud}
\bibinfo{author}{\bibfnamefont{L.}~\bibnamefont{Zou}},
  \bibinfo{author}{\bibfnamefont{H.~C.} \bibnamefont{Po}},
  \bibinfo{author}{\bibfnamefont{A.}~\bibnamefont{Vishwanath}},
  \bibnamefont{and} \bibinfo{author}{\bibfnamefont{T.}~\bibnamefont{Senthil}},
  \bibinfo{journal}{arXiv}
  (\bibinfo{year}{2018}), \eprint{1806.07873}.

\bibitem[{\citenamefont{Pizarro et~al.}(2018)\citenamefont{Pizarro,
  Calder{\'o}n, and Bascones}}]{Pizarro:2018wx}
\bibinfo{author}{\bibfnamefont{J.~M.} \bibnamefont{Pizarro}},
  \bibinfo{author}{\bibfnamefont{M.~J.} \bibnamefont{Calder{\'o}n}},
  \bibnamefont{and} \bibinfo{author}{\bibfnamefont{E.}~\bibnamefont{Bascones}},
  \bibinfo{journal}{arXiv}
  (\bibinfo{year}{2018}), \eprint{1805.07303}.

\bibitem[{\citenamefont{Isobe et~al.}(2018)\citenamefont{Isobe, Yuan, and
  Fu}}]{Isobe:2018wx}
\bibinfo{author}{\bibfnamefont{H.}~\bibnamefont{Isobe}},
  \bibinfo{author}{\bibfnamefont{N.~F.~Q.} \bibnamefont{Yuan}},
  \bibnamefont{and} \bibinfo{author}{\bibfnamefont{L.}~\bibnamefont{Fu}},
  \bibinfo{journal}{arXiv}
  (\bibinfo{year}{2018}), \eprint{1805.06449}.

\bibitem[{\citenamefont{Wu et~al.}(2018)\citenamefont{Wu, Pawlak, Jian, and
  Xu}}]{Wu:2018vh}
\bibinfo{author}{\bibfnamefont{X.-C.} \bibnamefont{Wu}},
  \bibinfo{author}{\bibfnamefont{K.~A.} \bibnamefont{Pawlak}},
  \bibinfo{author}{\bibfnamefont{C.-M.} \bibnamefont{Jian}}, \bibnamefont{and}
  \bibinfo{author}{\bibfnamefont{C.}~\bibnamefont{Xu}},
  \bibinfo{journal}{arXiv} (\bibinfo{year}{2018}),
  \eprint{1805.06906}.

\bibitem[{\citenamefont{You and Vishwanath}(2018)}]{You:2018ud}
\bibinfo{author}{\bibfnamefont{Y.-Z.} \bibnamefont{You}} \bibnamefont{and}
  \bibinfo{author}{\bibfnamefont{A.}~\bibnamefont{Vishwanath}},
  \bibinfo{journal}{arXiv}
  (\bibinfo{year}{2018}), \eprint{1805.06867}.

\bibitem[{\citenamefont{Sboychakov et~al.}(2018)\citenamefont{Sboychakov,
  Rozhkov, Rakhmanov, and Nori}}]{Sboychakov:2018tp}
\bibinfo{author}{\bibfnamefont{A.~O.} \bibnamefont{Sboychakov}},
  \bibinfo{author}{\bibfnamefont{A.~V.} \bibnamefont{Rozhkov}},
  \bibinfo{author}{\bibfnamefont{A.~L.} \bibnamefont{Rakhmanov}},
  \bibnamefont{and} \bibinfo{author}{\bibfnamefont{F.}~\bibnamefont{Nori}},
  \bibinfo{journal}{arXiv}
  (\bibinfo{year}{2018}), \eprint{1807.08190}.

\bibitem[{\citenamefont{Xu et~al.}(2018)\citenamefont{Xu, Law, and
  Lee}}]{Xu:2018vb}
\bibinfo{author}{\bibfnamefont{X.~Y.} \bibnamefont{Xu}},
  \bibinfo{author}{\bibfnamefont{K.~T.} \bibnamefont{Law}}, \bibnamefont{and}
  \bibinfo{author}{\bibfnamefont{P.~A.} \bibnamefont{Lee}},
  \bibinfo{journal}{arXiv}
  (\bibinfo{year}{2018}), \eprint{1805.00478}.

\bibitem[{\citenamefont{Ochi et~al.}(2018)\citenamefont{Ochi, Koshino, and
  Kuroki}}]{Ochi:2018ug}
\bibinfo{author}{\bibfnamefont{M.}~\bibnamefont{Ochi}},
  \bibinfo{author}{\bibfnamefont{M.}~\bibnamefont{Koshino}}, \bibnamefont{and}
  \bibinfo{author}{\bibfnamefont{K.}~\bibnamefont{Kuroki}},
  \bibinfo{journal}{arXiv}
  (\bibinfo{year}{2018}), \eprint{1805.09606}.

\bibitem[{\citenamefont{Sherkunov and Betouras}(2018)}]{Sherkunov:2018wf}
\bibinfo{author}{\bibfnamefont{Y.}~\bibnamefont{Sherkunov}} \bibnamefont{and}
  \bibinfo{author}{\bibfnamefont{J.~J.} \bibnamefont{Betouras}},
  \bibinfo{journal}{arXiv}
  (\bibinfo{year}{2018}), \eprint{1807.05524}.

\bibitem[{\citenamefont{Kennes et~al.}(2018)\citenamefont{Kennes, Lischner, and
  Karrasch}}]{Kennes:2018wi}
\bibinfo{author}{\bibfnamefont{D.~M.} \bibnamefont{Kennes}},
  \bibinfo{author}{\bibfnamefont{J.}~\bibnamefont{Lischner}}, \bibnamefont{and}
  \bibinfo{author}{\bibfnamefont{C.}~\bibnamefont{Karrasch}},
  \bibinfo{journal}{arXiv}
  (\bibinfo{year}{2018}), \eprint{1805.06310}.

\bibitem[{\citenamefont{Koshino et~al.}(2018)\citenamefont{Koshino, Yuan,
  Koretsune, Ochi, Kuroki, and Fu}}]{Koshino:2018td}
\bibinfo{author}{\bibfnamefont{M.}~\bibnamefont{Koshino}},
  \bibinfo{author}{\bibfnamefont{N.~F.~Q.} \bibnamefont{Yuan}},
  \bibinfo{author}{\bibfnamefont{T.}~\bibnamefont{Koretsune}},
  \bibinfo{author}{\bibfnamefont{M.}~\bibnamefont{Ochi}},
  \bibinfo{author}{\bibfnamefont{K.}~\bibnamefont{Kuroki}}, \bibnamefont{and}
  \bibinfo{author}{\bibfnamefont{L.}~\bibnamefont{Fu}},
  \bibinfo{journal}{arXiv} (\bibinfo{year}{2018}),
  \eprint{1805.06819}.

\bibitem[{\citenamefont{Kang and Vafek}(2018)}]{Kang:2018wd}
\bibinfo{author}{\bibfnamefont{J.}~\bibnamefont{Kang}} \bibnamefont{and}
  \bibinfo{author}{\bibfnamefont{O.}~\bibnamefont{Vafek}},
  \bibinfo{journal}{arXiv}
  (\bibinfo{year}{2018}), \eprint{1805.04918v2}.

\bibitem[{\citenamefont{Fidrysiak et~al.}(2018)\citenamefont{Fidrysiak,
  Zegrodnik, and Spalek}}]{Fidrysiak:2018}
\bibinfo{author}{\bibfnamefont{M.}~\bibnamefont{Fidrysiak}},
  \bibinfo{author}{\bibfnamefont{M.}~\bibnamefont{Zegrodnik}},
  \bibnamefont{and} \bibinfo{author}{\bibfnamefont{J.}~\bibnamefont{Spalek}},
  \bibinfo{journal}{arXiv}
  (\bibinfo{year}{2018}), \eprint{1805.01179v1}.

\bibitem[{\citenamefont{Song et~al.}(2018)\citenamefont{Song, Wang, Shi, Li,
  Fang, and Bernevig}}]{Song:2018ul}
\bibinfo{author}{\bibfnamefont{Z.}~\bibnamefont{Song}},
  \bibinfo{author}{\bibfnamefont{Z.}~\bibnamefont{Wang}},
  \bibinfo{author}{\bibfnamefont{W.}~\bibnamefont{Shi}},
  \bibinfo{author}{\bibfnamefont{G.}~\bibnamefont{Li}},
  \bibinfo{author}{\bibfnamefont{C.}~\bibnamefont{Fang}}, \bibnamefont{and}
  \bibinfo{author}{\bibfnamefont{B.~A.} \bibnamefont{Bernevig}},
  \bibinfo{journal}{arXiv}
  (\bibinfo{year}{2018}), \eprint{1807.10676}.

\bibitem[{\citenamefont{Venderbos and Fernandes}(2018)}]{Venderbos:2018vs}
\bibinfo{author}{\bibfnamefont{J.~W.~F.} \bibnamefont{Venderbos}}
  \bibnamefont{and} \bibinfo{author}{\bibfnamefont{R.~M.}
  \bibnamefont{Fernandes}}, 
  \bibinfo{journal}{arXiv} (\bibinfo{year}{2018}), \eprint{arXiv:1808.10416v1}.

\bibitem[{\citenamefont{Po et~al.}(2018{\natexlab{b}})\citenamefont{Po, Zou,
  Senthil, and Vishwanath}}]{Po:2018vk}
\bibinfo{author}{\bibfnamefont{H.~C.} \bibnamefont{Po}},
  \bibinfo{author}{\bibfnamefont{L.}~\bibnamefont{Zou}},
  \bibinfo{author}{\bibfnamefont{T.}~\bibnamefont{Senthil}}, \bibnamefont{and}
  \bibinfo{author}{\bibfnamefont{A.}~\bibnamefont{Vishwanath}},
  \bibinfo{journal}{arXiv}  (\bibinfo{year}{2018}{\natexlab{b}}),
  \eprint{1808.02482v2}.

\bibitem[{\citenamefont{Tang et~al.}(2018)\citenamefont{Tang, Yang, Wang,
  Zhang, and Wang}}]{Tang:2018tw}
\bibinfo{author}{\bibfnamefont{Q.~K.} \bibnamefont{Tang}},
  \bibinfo{author}{\bibfnamefont{L.}~\bibnamefont{Yang}},
  \bibinfo{author}{\bibfnamefont{D.}~\bibnamefont{Wang}},
  \bibinfo{author}{\bibfnamefont{F.~C.} \bibnamefont{Zhang}}, \bibnamefont{and}
  \bibinfo{author}{\bibfnamefont{Q.~H.} \bibnamefont{Wang}},
  \bibinfo{journal}{arXiv}  (\bibinfo{year}{2018}), \eprint{1809.06772v1}.

\bibitem[{\citenamefont{Laksono et~al.}(2018)\citenamefont{Laksono, Leaw,
  Reaves, Singh, Wang, Adam, and Gu}}]{Laksono:2018dm}
\bibinfo{author}{\bibfnamefont{E.}~\bibnamefont{Laksono}},
  \bibinfo{author}{\bibfnamefont{J.~N.} \bibnamefont{Leaw}},
  \bibinfo{author}{\bibfnamefont{A.}~\bibnamefont{Reaves}},
  \bibinfo{author}{\bibfnamefont{M.}~\bibnamefont{Singh}},
  \bibinfo{author}{\bibfnamefont{X.}~\bibnamefont{Wang}},
  \bibinfo{author}{\bibfnamefont{S.}~\bibnamefont{Adam}}, \bibnamefont{and}
  \bibinfo{author}{\bibfnamefont{X.}~\bibnamefont{Gu}},
  \bibinfo{journal}{arXiv} (\bibinfo{year}{2018}),
  \eprint{1808.04829v1}.

\bibitem[{\citenamefont{Tarnopolsky et~al.}(2018)\citenamefont{Tarnopolsky,
  Kruchkov, and Vishwanath}}]{Tarnopolsky:2018vv}
\bibinfo{author}{\bibfnamefont{G.}~\bibnamefont{Tarnopolsky}},
  \bibinfo{author}{\bibfnamefont{A.~J.} \bibnamefont{Kruchkov}},
  \bibnamefont{and}
  \bibinfo{author}{\bibfnamefont{A.}~\bibnamefont{Vishwanath}},
  \bibinfo{journal}{arXiv}  (\bibinfo{year}{2018}), \eprint{1808.05250v1}.

\bibitem[{\citenamefont{Chen et~al.}(2018)\citenamefont{Chen, Li, and
  Han}}]{Chen:2018vz}
\bibinfo{author}{\bibfnamefont{L.}~\bibnamefont{Chen}},
  \bibinfo{author}{\bibfnamefont{H.-Z.} \bibnamefont{Li}}, \bibnamefont{and}
  \bibinfo{author}{\bibfnamefont{R.-S.} \bibnamefont{Han}},
  \bibinfo{journal}{arXiv}  (\bibinfo{year}{2018}), \eprint{1809.00436v1}.

\bibitem[{\citenamefont{Lin and Nandkishore}(2018)}]{Lin:2018tx}
\bibinfo{author}{\bibfnamefont{Y.-P.} \bibnamefont{Lin}} \bibnamefont{and}
  \bibinfo{author}{\bibfnamefont{R.~M.} \bibnamefont{Nandkishore}},
  \bibinfo{journal}{arXiv}  (\bibinfo{year}{2018}), \eprint{1808.05270v1}.

\bibitem[{\citenamefont{Fu et~al.}(2018)\citenamefont{Fu, K{\"o}nig, Wilson,
  Chou, and Pixley}}]{Fu:2018vx}
\bibinfo{author}{\bibfnamefont{Y.}~\bibnamefont{Fu}},
  \bibinfo{author}{\bibfnamefont{E.~J.} \bibnamefont{K{\"o}nig}},
  \bibinfo{author}{\bibfnamefont{J.~H.} \bibnamefont{Wilson}},
  \bibinfo{author}{\bibfnamefont{Y.-Z.} \bibnamefont{Chou}}, \bibnamefont{and}
  \bibinfo{author}{\bibfnamefont{J.~H.} \bibnamefont{Pixley}},
  \bibinfo{journal}{arXiv}  (\bibinfo{year}{2018}), \eprint{1809.04604v1}.

\bibitem[{\citenamefont{Marzari et~al.}(2012)\citenamefont{Marzari, Mostofi,
  Yates, Souza, and Vanderbilt}}]{Marzari:2012eu}
\bibinfo{author}{\bibfnamefont{N.}~\bibnamefont{Marzari}},
  \bibinfo{author}{\bibfnamefont{A.~A.} \bibnamefont{Mostofi}},
  \bibinfo{author}{\bibfnamefont{J.~R.} \bibnamefont{Yates}},
  \bibinfo{author}{\bibfnamefont{I.}~\bibnamefont{Souza}}, \bibnamefont{and}
  \bibinfo{author}{\bibfnamefont{D.}~\bibnamefont{Vanderbilt}},
  \bibinfo{journal}{Rev. Mod. Phys.} \textbf{\bibinfo{volume}{84}},
  \bibinfo{pages}{1419} (\bibinfo{year}{2012}).

\bibitem[{\citenamefont{Marzari and Vanderbilt}(1997)}]{Marzari:1997wa}
\bibinfo{author}{\bibfnamefont{N.}~\bibnamefont{Marzari}} \bibnamefont{and}
  \bibinfo{author}{\bibfnamefont{D.}~\bibnamefont{Vanderbilt}},
  \bibinfo{journal}{Phys. Rev. B} \textbf{\bibinfo{volume}{56}},
  \bibinfo{pages}{12847} (\bibinfo{year}{1997}).

\bibitem[{\citenamefont{Wong et~al.}(2015)\citenamefont{Wong, Wang, Jung,
  Pezzini, DaSilva, Tsai, Jung, Khajeh, Kim, Lee et~al.}}]{Wong:2015dg}
\bibinfo{author}{\bibfnamefont{D.}~\bibnamefont{Wong}},
  \bibinfo{author}{\bibfnamefont{Y.}~\bibnamefont{Wang}},
  \bibinfo{author}{\bibfnamefont{J.}~\bibnamefont{Jung}},
  \bibinfo{author}{\bibfnamefont{S.}~\bibnamefont{Pezzini}},
  \bibinfo{author}{\bibfnamefont{A.~M.} \bibnamefont{DaSilva}},
  \bibinfo{author}{\bibfnamefont{H.-Z.} \bibnamefont{Tsai}},
  \bibinfo{author}{\bibfnamefont{H.~S.} \bibnamefont{Jung}},
  \bibinfo{author}{\bibfnamefont{R.}~\bibnamefont{Khajeh}},
  \bibinfo{author}{\bibfnamefont{Y.}~\bibnamefont{Kim}},
  \bibinfo{author}{\bibfnamefont{J.}~\bibnamefont{Lee}}, \bibnamefont{et~al.},
  \bibinfo{journal}{Phys. Rev. B} \textbf{\bibinfo{volume}{92}},
  \bibinfo{pages}{155409} (\bibinfo{year}{2015}).

\bibitem[{\citenamefont{Reich et~al.}(2002)\citenamefont{Reich, Maultzsch,
  Thomsen, and Ordej{\'o}n}}]{Reich:2002kd}
\bibinfo{author}{\bibfnamefont{S.}~\bibnamefont{Reich}},
  \bibinfo{author}{\bibfnamefont{J.}~\bibnamefont{Maultzsch}},
  \bibinfo{author}{\bibfnamefont{C.}~\bibnamefont{Thomsen}}, \bibnamefont{and}
  \bibinfo{author}{\bibfnamefont{P.}~\bibnamefont{Ordej{\'o}n}},
  \bibinfo{journal}{Phys. Rev. B} \textbf{\bibinfo{volume}{66}},
  \bibinfo{pages}{035412} (\bibinfo{year}{2002}).

\bibitem[{\citenamefont{Castro~Neto et~al.}(2009)\citenamefont{Castro~Neto,
  Guinea, Peres, Novoselov, and Geim}}]{CastroNeto:2009cl}
\bibinfo{author}{\bibfnamefont{A.~H.} \bibnamefont{Castro~Neto}},
  \bibinfo{author}{\bibfnamefont{F.}~\bibnamefont{Guinea}},
  \bibinfo{author}{\bibfnamefont{N.~M.~R.} \bibnamefont{Peres}},
  \bibinfo{author}{\bibfnamefont{K.~S.} \bibnamefont{Novoselov}},
  \bibnamefont{and} \bibinfo{author}{\bibfnamefont{A.~K.} \bibnamefont{Geim}},
  \bibinfo{journal}{Rev. Mod. Phys.} \textbf{\bibinfo{volume}{81}},
  \bibinfo{pages}{109} (\bibinfo{year}{2009}).

\bibitem[{\citenamefont{Su{\'a}rez~Morell
  et~al.}(2010)\citenamefont{Su{\'a}rez~Morell, Correa, Vargas, Pacheco, and
  Barticevic}}]{SuarezMorell:2010bz}
\bibinfo{author}{\bibfnamefont{E.}~\bibnamefont{Su{\'a}rez~Morell}},
  \bibinfo{author}{\bibfnamefont{J.~D.} \bibnamefont{Correa}},
  \bibinfo{author}{\bibfnamefont{P.}~\bibnamefont{Vargas}},
  \bibinfo{author}{\bibfnamefont{M.}~\bibnamefont{Pacheco}}, \bibnamefont{and}
  \bibinfo{author}{\bibfnamefont{Z.}~\bibnamefont{Barticevic}},
  \bibinfo{journal}{Phys. Rev. B} \textbf{\bibinfo{volume}{82}},
  \bibinfo{pages}{121407} (\bibinfo{year}{2010}).

\bibitem[{\citenamefont{Dresselhaus et~al.}(2007)\citenamefont{Dresselhaus,
  Dresselhaus, and Jorio}}]{Dresselhaus:2007wd}
\bibinfo{author}{\bibfnamefont{M.~S.} \bibnamefont{Dresselhaus}},
  \bibinfo{author}{\bibfnamefont{G.}~\bibnamefont{Dresselhaus}},
  \bibnamefont{and} \bibinfo{author}{\bibfnamefont{A.}~\bibnamefont{Jorio}},
  \emph{\bibinfo{title}{{Group theory: application to the physics of condensed
  matter}}}, \bibinfo{howpublished}{Springer} (\bibinfo{year}{2007}).

\bibitem[{\citenamefont{Bradlyn et~al.}(2017)\citenamefont{Bradlyn, Elcoro,
  Cano, Vergniory, Wang, Felser, Aroyo, and Bernevig}}]{Bradlyn:2017fy}
\bibinfo{author}{\bibfnamefont{B.}~\bibnamefont{Bradlyn}},
  \bibinfo{author}{\bibfnamefont{L.}~\bibnamefont{Elcoro}},
  \bibinfo{author}{\bibfnamefont{J.}~\bibnamefont{Cano}},
  \bibinfo{author}{\bibfnamefont{M.~G.} \bibnamefont{Vergniory}},
  \bibinfo{author}{\bibfnamefont{Z.}~\bibnamefont{Wang}},
  \bibinfo{author}{\bibfnamefont{C.}~\bibnamefont{Felser}},
  \bibinfo{author}{\bibfnamefont{M.~I.} \bibnamefont{Aroyo}}, \bibnamefont{and}
  \bibinfo{author}{\bibfnamefont{B.~A.} \bibnamefont{Bernevig}},
  \bibinfo{journal}{Nature} \textbf{\bibinfo{volume}{547}},
  \bibinfo{pages}{298} (\bibinfo{year}{2017}).

\bibitem[{\citenamefont{Mele}(2011)}]{Mele:2011cb}
\bibinfo{author}{\bibfnamefont{E.~J.} \bibnamefont{Mele}},
  \bibinfo{journal}{Phys. Rev. B} \textbf{\bibinfo{volume}{84}},
  \bibinfo{pages}{235439} (\bibinfo{year}{2011}).

\bibitem[{\citenamefont{Fang and Kaxiras}(2016)}]{Fang:2016iq}
\bibinfo{author}{\bibfnamefont{S.}~\bibnamefont{Fang}} \bibnamefont{and}
  \bibinfo{author}{\bibfnamefont{E.}~\bibnamefont{Kaxiras}},
  \bibinfo{journal}{Phys. Rev. B} \textbf{\bibinfo{volume}{93}},
  \bibinfo{pages}{235153} (\bibinfo{year}{2016}).

\bibitem[{\citenamefont{Jung and MacDonald}(2014)}]{Jung:2014hj}
\bibinfo{author}{\bibfnamefont{J.}~\bibnamefont{Jung}} \bibnamefont{and}
  \bibinfo{author}{\bibfnamefont{A.~H.} \bibnamefont{MacDonald}},
  \bibinfo{journal}{Phys. Rev. B} \textbf{\bibinfo{volume}{89}},
  \bibinfo{pages}{035405} (\bibinfo{year}{2014}).

\bibitem[{\citenamefont{Malard et~al.}(2007)\citenamefont{Malard, Nilsson,
  Elias, Brant, Plentz, Alves, Castro~Neto, and Pimenta}}]{Malard:2007jv}
\bibinfo{author}{\bibfnamefont{L.~M.} \bibnamefont{Malard}},
  \bibinfo{author}{\bibfnamefont{J.}~\bibnamefont{Nilsson}},
  \bibinfo{author}{\bibfnamefont{D.~C.} \bibnamefont{Elias}},
  \bibinfo{author}{\bibfnamefont{J.~C.} \bibnamefont{Brant}},
  \bibinfo{author}{\bibfnamefont{F.}~\bibnamefont{Plentz}},
  \bibinfo{author}{\bibfnamefont{E.~S.} \bibnamefont{Alves}},
  \bibinfo{author}{\bibfnamefont{A.~H.} \bibnamefont{Castro~Neto}},
  \bibnamefont{and} \bibinfo{author}{\bibfnamefont{M.~A.}
  \bibnamefont{Pimenta}}, \bibinfo{journal}{Phys. Rev. B}
  \textbf{\bibinfo{volume}{76}}, \bibinfo{pages}{201401}
  (\bibinfo{year}{2007}).

\bibitem[{\citenamefont{Bistritzer and MacDonald}(2011)}]{Bistritzer:2011ho}
\bibinfo{author}{\bibfnamefont{R.}~\bibnamefont{Bistritzer}} \bibnamefont{and}
  \bibinfo{author}{\bibfnamefont{A.~H.} \bibnamefont{MacDonald}},
  \bibinfo{journal}{Proc Natl Acad Sci USA} \textbf{\bibinfo{volume}{108}},
  \bibinfo{pages}{12233} (\bibinfo{year}{2011}).

\bibitem[{\citenamefont{Lopes~dos Santos et~al.}(2012)\citenamefont{Lopes~dos
  Santos, Peres, and Castro~Neto}}]{LopesdosSantos:2012vk}
\bibinfo{author}{\bibfnamefont{J.~M.~B.} \bibnamefont{Lopes~dos Santos}},
  \bibinfo{author}{\bibfnamefont{N.~M.~R.}~\bibnamefont{Peres}}, \bibnamefont{and}
  \bibinfo{author}{\bibfnamefont{A.~H.} \bibnamefont{Castro~Neto}},
  \bibinfo{journal}{Phys. Rev. B} \textbf{\bibinfo{volume}{86}},
  \bibinfo{pages}{155449} (\bibinfo{year}{2012}).

\bibitem[{\citenamefont{Choi and Choi}(2018)}]{Choi:2018vw}
\bibinfo{author}{\bibfnamefont{Y.~W.} \bibnamefont{Choi}} \bibnamefont{and}
  \bibinfo{author}{\bibfnamefont{H.~J.} \bibnamefont{Choi}},
  \bibinfo{journal}{arXiv}  (\bibinfo{year}{2018}), \eprint{1809.08407v1}.

\bibitem[{\citenamefont{Angeli et~al.}(2018)\citenamefont{Angeli, Mandelli,
  Valli, Amaricci, Capone, Tosatti, and Fabrizio}}]{Angeli:2018wy}
\bibinfo{author}{\bibfnamefont{M.}~\bibnamefont{Angeli}},
  \bibinfo{author}{\bibfnamefont{D.}~\bibnamefont{Mandelli}},
  \bibinfo{author}{\bibfnamefont{A.}~\bibnamefont{Valli}},
  \bibinfo{author}{\bibfnamefont{A.}~\bibnamefont{Amaricci}},
  \bibinfo{author}{\bibfnamefont{M.}~\bibnamefont{Capone}},
  \bibinfo{author}{\bibfnamefont{E.}~\bibnamefont{Tosatti}}, \bibnamefont{and}
  \bibinfo{author}{\bibfnamefont{M.}~\bibnamefont{Fabrizio}},
  \bibinfo{journal}{arXiv}  (\bibinfo{year}{2018}), \eprint{1809.11140v1}.

\bibitem[{\citenamefont{Rademaker et~al.}(2013)\citenamefont{Rademaker,
  Pramudya, Zaanen, and Dobrosavljevi{\'c}}}]{Rademaker:2013jg}
\bibinfo{author}{\bibfnamefont{L.}~\bibnamefont{Rademaker}},
  \bibinfo{author}{\bibfnamefont{Y.}~\bibnamefont{Pramudya}},
  \bibinfo{author}{\bibfnamefont{J.}~\bibnamefont{Zaanen}}, \bibnamefont{and}
  \bibinfo{author}{\bibfnamefont{V.}~\bibnamefont{Dobrosavljevi{\'c}}},
  \bibinfo{journal}{Phys. Rev. E} \textbf{\bibinfo{volume}{88}},
  \bibinfo{pages}{032121} (\bibinfo{year}{2013}).

\bibitem[{\citenamefont{Schuler et~al.}(2013)\citenamefont{Schuler, Rosner,
  Wehling, Lichtenstein, and Katsnelson}}]{Schuler:2013ip}
\bibinfo{author}{\bibfnamefont{M.}~\bibnamefont{Schuler}},
  \bibinfo{author}{\bibfnamefont{M.}~\bibnamefont{Rosner}},
  \bibinfo{author}{\bibfnamefont{T.~O.} \bibnamefont{Wehling}},
  \bibinfo{author}{\bibfnamefont{A.~I.} \bibnamefont{Lichtenstein}},
  \bibnamefont{and} \bibinfo{author}{\bibfnamefont{M.~I.}
  \bibnamefont{Katsnelson}}, \bibinfo{journal}{Phys. Rev. Lett.}
  \textbf{\bibinfo{volume}{111}}, \bibinfo{pages}{036601}
  (\bibinfo{year}{2013}).

\bibitem[{\citenamefont{Wehling et~al.}(2011)\citenamefont{Wehling, Sasioglu,
  Friedrich, Lichtenstein, Katsnelson, and Blugel}}]{Wehling:2011cf}
\bibinfo{author}{\bibfnamefont{T.~O.} \bibnamefont{Wehling}},
  \bibinfo{author}{\bibfnamefont{E.}~\bibnamefont{Sasioglu}},
  \bibinfo{author}{\bibfnamefont{C.}~\bibnamefont{Friedrich}},
  \bibinfo{author}{\bibfnamefont{A.~I.} \bibnamefont{Lichtenstein}},
  \bibinfo{author}{\bibfnamefont{M.~I.} \bibnamefont{Katsnelson}},
  \bibnamefont{and} \bibinfo{author}{\bibfnamefont{S.}~\bibnamefont{Blugel}},
  \bibinfo{journal}{Phys. Rev. Lett.} \textbf{\bibinfo{volume}{106}},
  \bibinfo{pages}{236805} (\bibinfo{year}{2011}).

\bibitem[{\citenamefont{Zaanen et~al.}(1985)\citenamefont{Zaanen, Sawatzky, and
  Allen}}]{Zaanen:1985bsa}
\bibinfo{author}{\bibfnamefont{J.}~\bibnamefont{Zaanen}},
  \bibinfo{author}{\bibfnamefont{G.~A.} \bibnamefont{Sawatzky}},
  \bibnamefont{and} \bibinfo{author}{\bibfnamefont{J.~W.} \bibnamefont{Allen}},
  \bibinfo{journal}{Phys. Rev. Lett.} \textbf{\bibinfo{volume}{55}},
  \bibinfo{pages}{418} (\bibinfo{year}{1985}).

\bibitem[{\citenamefont{Imada et~al.}(1998)\citenamefont{Imada, Fujimori, and
  Tokura}}]{Imada:1998er}
\bibinfo{author}{\bibfnamefont{M.}~\bibnamefont{Imada}},
  \bibinfo{author}{\bibfnamefont{A.}~\bibnamefont{Fujimori}}, \bibnamefont{and}
  \bibinfo{author}{\bibfnamefont{Y.}~\bibnamefont{Tokura}},
  \bibinfo{journal}{Rev. Mod. Phys.} \textbf{\bibinfo{volume}{70}},
  \bibinfo{pages}{1039} (\bibinfo{year}{1998}).

\bibitem[{\citenamefont{Georges et~al.}(2013)\citenamefont{Georges, Medici, and
  Mravlje}}]{Georges:2013ju}
\bibinfo{author}{\bibfnamefont{A.}~\bibnamefont{Georges}},
  \bibinfo{author}{\bibfnamefont{L.~d.} \bibnamefont{Medici}},
  \bibnamefont{and} \bibinfo{author}{\bibfnamefont{J.}~\bibnamefont{Mravlje}},
  \bibinfo{journal}{Annu. Rev. Condens. Matter Phys.}
  \textbf{\bibinfo{volume}{4}}, \bibinfo{pages}{137} (\bibinfo{year}{2013}).

\bibitem[{\citenamefont{Hastings}(2004)}]{Hastings:2004cd}
\bibinfo{author}{\bibfnamefont{M.~B.} \bibnamefont{Hastings}},
  \bibinfo{journal}{Phys. Rev. B} \textbf{\bibinfo{volume}{69}},
  \bibinfo{pages}{104431} (\bibinfo{year}{2004}).

\bibitem[{\citenamefont{Spa{\l}ek}(1988)}]{Spaiek:1988cz}
\bibinfo{author}{\bibfnamefont{J.}~\bibnamefont{Spa{\l}ek}},
  \bibinfo{journal}{Phys. Rev. B} \textbf{\bibinfo{volume}{37}},
  \bibinfo{pages}{533} (\bibinfo{year}{1988}).

\bibitem[{\citenamefont{Klein and Seitz}(1973)}]{1973PhRvB...8.2236K}
\bibinfo{author}{\bibfnamefont{D.~J.} \bibnamefont{Klein}} \bibnamefont{and}
  \bibinfo{author}{\bibfnamefont{W.~A.} \bibnamefont{Seitz}},
  \bibinfo{journal}{Phys. Rev. B} \textbf{\bibinfo{volume}{8}},
  \bibinfo{pages}{2236} (\bibinfo{year}{1973}).

\bibitem[{\citenamefont{Chen et~al.}(2013)\citenamefont{Chen, Meng, Yu, Yang,
  Jarrell, and Moreno}}]{Chen:2013hg}
\bibinfo{author}{\bibfnamefont{K.~S.} \bibnamefont{Chen}},
  \bibinfo{author}{\bibfnamefont{Z.~Y.} \bibnamefont{Meng}},
  \bibinfo{author}{\bibfnamefont{U.}~\bibnamefont{Yu}},
  \bibinfo{author}{\bibfnamefont{S.}~\bibnamefont{Yang}},
  \bibinfo{author}{\bibfnamefont{M.}~\bibnamefont{Jarrell}}, \bibnamefont{and}
  \bibinfo{author}{\bibfnamefont{J.}~\bibnamefont{Moreno}},
  \bibinfo{journal}{Phys. Rev. B} \textbf{\bibinfo{volume}{88}},
  \bibinfo{pages}{041103} (\bibinfo{year}{2013}).

\bibitem[{\citenamefont{Uchoa and Castro~Neto}(2007)}]{Uchoa:2007bq}
\bibinfo{author}{\bibfnamefont{B.}~\bibnamefont{Uchoa}} \bibnamefont{and}
  \bibinfo{author}{\bibfnamefont{A.~H.} \bibnamefont{Castro~Neto}},
  \bibinfo{journal}{Phys. Rev. Lett.} \textbf{\bibinfo{volume}{98}},
  \bibinfo{pages}{146801} (\bibinfo{year}{2007}).

\bibitem[{\citenamefont{Slotman et~al.}(2015)\citenamefont{Slotman, van Wijk,
  Zhao, Fasolino, Katsnelson, and Yuan}}]{Slotman:2015kd}
\bibinfo{author}{\bibfnamefont{G.~J.} \bibnamefont{Slotman}},
  \bibinfo{author}{\bibfnamefont{M.~M.} \bibnamefont{van Wijk}},
  \bibinfo{author}{\bibfnamefont{P.-L.} \bibnamefont{Zhao}},
  \bibinfo{author}{\bibfnamefont{A.}~\bibnamefont{Fasolino}},
  \bibinfo{author}{\bibfnamefont{M.~I.} \bibnamefont{Katsnelson}},
  \bibnamefont{and} \bibinfo{author}{\bibfnamefont{S.}~\bibnamefont{Yuan}},
  \bibinfo{journal}{Phys. Rev. Lett.} \textbf{\bibinfo{volume}{115}},
  \bibinfo{pages}{186801} (\bibinfo{year}{2015}).

\bibitem[{\citenamefont{van Wijk et~al.}(2015)\citenamefont{van Wijk, Schuring,
  Katsnelson, and Fasolino}}]{vanWijk:2015bc}
\bibinfo{author}{\bibfnamefont{M.~M.} \bibnamefont{van Wijk}},
  \bibinfo{author}{\bibfnamefont{A.}~\bibnamefont{Schuring}},
  \bibinfo{author}{\bibfnamefont{M.~I.} \bibnamefont{Katsnelson}},
  \bibnamefont{and} \bibinfo{author}{\bibfnamefont{A.}~\bibnamefont{Fasolino}},
  \bibinfo{journal}{2D Materials} \textbf{\bibinfo{volume}{2}},
  \bibinfo{pages}{034010} (\bibinfo{year}{2015}).

\bibitem[{\citenamefont{Nam and Koshino}(2017)}]{Nam:2017jh}
\bibinfo{author}{\bibfnamefont{N.~N.~T.} \bibnamefont{Nam}} \bibnamefont{and}
  \bibinfo{author}{\bibfnamefont{M.}~\bibnamefont{Koshino}},
  \bibinfo{journal}{Phys. Rev. B} \textbf{\bibinfo{volume}{96}},
  \bibinfo{pages}{075311} (\bibinfo{year}{2017}).

\bibitem[{\citenamefont{Ament et~al.}(2011)\citenamefont{Ament, van Veenendaal,
  Devereaux, Hill, and van~den Brink}}]{Ament:2011jy}
\bibinfo{author}{\bibfnamefont{L.~J.~P.} \bibnamefont{Ament}},
  \bibinfo{author}{\bibfnamefont{M.}~\bibnamefont{van Veenendaal}},
  \bibinfo{author}{\bibfnamefont{T.~P.} \bibnamefont{Devereaux}},
  \bibinfo{author}{\bibfnamefont{J.~P.} \bibnamefont{Hill}}, \bibnamefont{and}
  \bibinfo{author}{\bibfnamefont{J.}~\bibnamefont{van~den Brink}},
  \bibinfo{journal}{Rev. Mod. Phys.} \textbf{\bibinfo{volume}{83}},
  \bibinfo{pages}{705} (\bibinfo{year}{2011}).

\bibitem[{\citenamefont{Van~Harlingen}(1995)}]{VanHarlingen:1995ht}
\bibinfo{author}{\bibfnamefont{D.~J.} \bibnamefont{Van~Harlingen}},
  \bibinfo{journal}{Rev. Mod. Phys.} \textbf{\bibinfo{volume}{67}},
  \bibinfo{pages}{515} (\bibinfo{year}{1995}).

\end{thebibliography}
\end{document}